\documentclass[fleqn,usenatbib]{rasti}

\usepackage{newtxtext,newtxmath}
\usepackage{float}
\usepackage{multirow}
\usepackage{longtable}
\usepackage[T1]{fontenc}
\usepackage[version=4]{mhchem}
\usepackage{array}
\usepackage{geometry}
\usepackage{ragged2e}
\usepackage{hyperref}
\usepackage{comment}
\usepackage{makecell} 
 \usepackage{color}
\usepackage{amsmath} 
\usepackage{booktabs}
\usepackage{graphicx}
\usepackage{makecell}
\usepackage{xcolor}
\usepackage{listings}

\DeclareRobustCommand{\VAN}[3]{#2}
\let\VANthebibliography\thebibliography
\def\thebibliography{\DeclareRobustCommand{\VAN}[3]{##3}\VANthebibliography}
\definecolor{pink}{RGB}{255, 105, 180} 

\usepackage{graphicx}	
\usepackage{amsmath}	
\usepackage{xcolor}
\newcolumntype{P}[1]{>{\raggedright\arraybackslash}p{#1}}

\lstdefinelanguage{json}{
    basicstyle=\ttfamily\small,       
    numbers=left,                     
    numberstyle=\tiny\color{gray},    
    stepnumber=1,                     
    numbersep=5pt,                    
    showstringspaces=false,           
    breaklines=true,                  
    frame=single,                     
    backgroundcolor=\color{gray!10},  
    literate=
     *{0}{{{\color{blue}0}}}{1}       
      {1}{{{\color{blue}1}}}{1}
      {2}{{{\color{blue}2}}}{1}
      {3}{{{\color{blue}3}}}{1}
      {4}{{{\color{blue}4}}}{1}
      {5}{{{\color{blue}5}}}{1}
      {6}{{{\color{blue}6}}}{1}
      {7}{{{\color{blue}7}}}{1}
      {8}{{{\color{blue}8}}}{1}
      {9}{{{\color{blue}9}}}{1}
      {:}{{{\color{red}{:}}}}{1}      
      {,}{{{\color{red}{,}}}}{1}      
       {"}{{{\color{green!50!black}{"}}}}{1},     
}

\title[The \textsc{ExoPhoto} Database]{\textsc{ExoPhoto}: A Database of Temperature-Dependent Photodissociation Cross Sections}

\author[Q.-H. Ni et al.]{Qing-He Ni,$^{1}$
Christian Hill,$^{2}$
Sergei N. Yurchenko,$^{1}$
Marco Pezzella,$^{3}$
Alexander Fateev,$^{4}$
Zhi Qin,$^{5}$
\newauthor{Olivia Venot,$^{6}$
and Jonathan Tennyson$^{1}$ \thanks{E-mail: j.tennyson@ucl.ac.uk}}
\\
$^{1}$Department of Physics and Astronomy, University College London, Gower Street, London WC1E 6BT, UK\\
$^{2}$International Atomic Energy Agency, Vienna A-1400, Austria\\
$^{3}$Dipartimento di Fisica e Geologia,  Universit\'{a} di Perugia, Via Alessandro Pascoli, Perugia, Italy \\
$^{4}$Department of Chemical and Biochemical Engineering, Technical University of Denmark, S\o ltofts Plads 229, Kgs. Lyngby, DK 2800, Denmark \\
$^{5}$School of Energy and Power Engineering, Shandong University, Jinan 250061, China \\
$^{6}$Universit\'{e} Paris Cit\'{e} and Univ Paris Est Creteil, CNRS, LISA, F-75013 Paris, France \\
}

\date{Accepted XXX. Received YYY; in original form ZZZ}

\pubyear{\the\year{}}

\begin{document}
\label{firstpage}
\pagerange{\pageref{firstpage}--\pageref{lastpage}}
\maketitle

\begin{abstract}

We present the \textsc{ExoPhoto} database  \url{https://exomol.com/exophoto/}, an extension of the ExoMol database, specifically developed to address the growing need for high-accuracy, temperature-dependent photodissociation cross section data towards short-UV wavelengths. \textsc{ExoPhoto} combines theoretical models from three major computational databases (\textsc{ExoMol}, \textsc{UGAMOP} and \textsc{PhoMol}) and experimental datasets from two experimental groups, providing extensive wavelength and temperature coverage.
\textsc{ExoPhoto} currently includes photodissociation data for 20 molecules: \( \text{AlH} \), \( \text{HCl} \), \( \text{HF} \), \( \text{MgH} \), \( \text{OH} \), \( \text{NaO} \), \( \text{MgO} \), \( \text{O}_2 \), \( \text{AlCl} \), \( \text{AlF} \), \( \text{CS} \), \( \text{HeH}^+ \), \( \text{CO} \), \( \text{CO}_2 \), \( \text{H}_2\text{O} \), \( \text{SO}_2 \), \( \text{C}_2\text{H}_2 \), \( \text{C}_2\text{H}_4 \), \( \text{H}_2\text{CO} \), and \( \text{NH}_3 \), derived from theoretical models and supported by experimental data from 5 databases.

\textsc{ExoPhoto} also includes detailed data on branching ratios and quantum yields for selected datasets. The data structure of \textsc{ExoPhoto} follows the \textsc{ExoMol} framework, with a consistent naming convention and hierarchical JSON-based organization. Photodissociation cross sections are stored in a set of \texttt{.photo} files which provide data as a function of wavelength with one file for each target molecule temperature.
Future developments aim to include more photodissociation cross section data and to provide data for molecules in non-local thermodynamic equilibrium (non-LTE). 
These will expand the utility of \textsc{ExoPhoto} for advanced astrophysical, planetary modeling and industrial applications.

\end{abstract}



\onecolumn

\section{Introduction}



Photodissociation is the process in which chemical bonds are broken due to the absorption of one or more photons. This is a very important process in astronomy because the chemistry of upper planetary atmospheres, as in case of the Earth’s atmosphere, and the intersteller medium is heavily reliant on photodissociation. Molecular temperature can have a significant impact on the photodissociation rate since rising temperatures increase the number of vibrationally excited states populated and therefore decrease the photodissociation threshold \citep{jt229}. As a result, the photodissociation rates in exoplanets orbiting close to cooler stars can rise rapidly at molecular temperatures beyond 1000~K, but the details of this behaviour depends strongly on the  radiation fields \citep{93Schinke,jt865}. An important aspect in the evolution of the atmospheres of exoplanets is photon-initiated chemistry. In this phenomenon, the composition and dynamics of the atmosphere for planets orbiting stars in ultraviolet (UV) rich environments caused by high-energy photons are dominated by photodissociation \citep{intro1,intro2,intro3,intro4}.

In recent years, photodissociation databases have become  a critical resource for exploring molecular interactions under UV radiation, with applications spanning planetary atmospheres to interstellar environments. The  \href{https://home.strw.leidenuniv.nl/~ewine/photo/}{Leiden} Database \citep{leiden}, established in the 1980s, has served as a cornerstone for astrochemical modeling, particularly for studies involving interstellar media (ISM) and photon-dominated regions (PDRs). This database compiles the essential molecular data required to calculate photodestruction rates in various astrophysical contexts. It is constructed based on laboratory and theoretical datasets, ensuring the inclusion of species with verified cross section data. Currently, it features cross sections for 116 atoms and molecules covering photoabsorption, photonionisation and, for molecules, photodissociation,  with no temperature dependence indicated and a tacit assumptions that the molecules are cold.


The \href{https://www.uv-vis-spectral-atlas-mainz.org/uvvis/}{MPI-Mainz UV/VIS Spectral Atlas} \citep{MPIMainz}, initially developed in the early 1980s and made available online in 2003, provides one of the most extensive datasets for atmospheric research, covering nearly a century of mainly experimental studies. With data for approximately 900 species, it includes UV and visible absorption cross sections and quantum yields, totaling over 5500 cross section files. While it offers extensive UV and visible absorption cross sections,  the data provided is largely recorded at room temperature.  These resources are vital for understanding kinetic and photochemical processes in atmospheric chemistry.  Building upon this, the \href{http://www.vamdc-project.vamdc.eu/}{VAMDC (Virtual Atomic and Molecular Data Centre) } \citep{VAMDC,jt814},  integrates the \textsc{MPI-Mainz} database within its unified e-science framework, enabling seamless access to photodissociation data from multiple sources through a single interface.

The \href{https://sites.physast.uga.edu/ugamop/}{UGAMOP} Database (University of Georgia Molecular Opacity Project) \citep{UGAMOP} was designed to address astrophysical needs for accurate line and continuum opacity data. It features calculated temperature-dependent photodissociation cross sections for four molecules: CS \citep{ugaCS}, H$_2$ \citep{ugah2}, HeH$^+$ \citep{ugaHeH}, and MgH \citep{ugamgh1,ugamgh2,ugamgh3}. The photodissociation cross sections for CS, H$_2$, HeH$^+$, and MgH, all assume local thermodyamic equilibrium (LTE), while for  H$_2$  non-LTE is also allowed but not all photodissociaiton pathways are considered. These datasets are particularly significant for modeling extrasolar giant planets (EGPs) and cool stellar atmospheres.

In 2014, \href{https://phidrates.space.swri.edu/}{PHIDRATES} (PhotoIonization and Dissociation Rates for Atmospheric and Terrestrial Environments) \citep{phidrate} was introduced, providing detailed photoionization and photodissociation rates. This database supports studies involving solar and blackbody radiation fields and is particularly useful for comet modeling, planetary atmospheres, and heliospheric dynamics. Rate coefficients for ionization and dissociation have been determined for more than 140 atoms, molecules, and ions within radiation fields generated by both the quiet and active Sun at a heliocentric distance of 1~AU, as well as from black bodies characterized by selected temperatures spanning from 1000~K to $10^6$~K, without considering any radiation dilution due to the distance from these sources or the  dependence on the temperature of molecular species.

The \href{http://servo.aob.rs/mold/}{MOL-D} Database \citep{MOLD}, hosted by the Belgrade Astronomical Observatory since 2015, focuses on temperature-independent photodissociation cross sections for ro-vibrational states of molecular ions, including  \(\mathrm{He}_2^+\), \(\mathrm{H}_2^+\), and \(\mathrm{LiH}^+\). As part of the Serbian Virtual Observatory \textsc{(SerVO)} and \textsc{VAMDC}, this database offers cross section data and rate coefficients for specific collisional processes, accessible via a user-friendly web interface. 

Since 2017, the Metal Oxides Photolysis database \citep{metaloxide}, has provided quantum chemistry-based photolysis cross sections for molecules like LiO, NaO, and MgO focusing on high-temperature environments, although the cross sections provided are not actually temperature-dependent. These datasets are crucial for studying planetary exospheres, as photolysis of metal oxides contributes to the release of metal atoms, impacting observations of exospheric dynamics. 

Recently,  a photodissociation cross section project called \textsc{PhoMol}  has been initiated by a research team at the School of Energy and Power Engineering, Shandong University in China which has computed temperature-dependent cross sections for a range of astrophysically important diatomic molecules, including, AlCl \citep{qinAlCl}, AlF \citep{qinAlF}, AlH \citep{qinAlh}, HCl \citep{qinHclHF}, HF \citep{qinHclHF}, MgO \citep{qinMgO}, NaO \citep{qinNaO}, and O$_2$ \citep{qinO2}. These photodissociation cross sections are considered further below.

Together, these databases and datasets form a network of resources, enabling simulations of molecular behavior in diverse radiation environments and advancing our understanding of astrophysical, atmospheric, and industrial photochemistry.

The \textsc{ExoMol} project was originally established to provide
extensive, temperature-dependent spectroscopic data sets involving bound-bound transitions \citep{jt528}. However, modellers also need information on dissociative processes as this often drives chemistry. 

To address this issue, the \textsc{ExoMol} project, is undertaking a significant expansion of its  database (\url{www.exomol.com}) to consider processes at ultraviolet wavelengths including continuum absorption \citep{jt922}, predissociation \citep{jt922,jt933,jt969} and photodissociation \citep{jt840,jt865,jt958}.  At present, the \textsc{ExoMol} database is largely a molecular line lists database that can be used to simulate photo-absorption and radiative transfer in astrophysical environments such as exoplanets, brown dwarfs, cold stars, and sunspots and characterize their spectral properties. Extensive line lists are the basis of the database \citep{jt939}; they are then complemented with state lifetimes, dipoles, Land\'{e} $g$-factors, cooling functions, temperature-dependent cross sections, opacity functions, specific heats etc. Due to their completeness as a function of frequency and their  coverage of molecular species, and, most critically, temperature, \textsc{ExoMol} data are widely used by astronomers studying exoplanet and other objects.

The \textsc{ExoPhoto} Database extends the \textsc{ExoMol} database by providing photodissociation cross section data extracted from \textsc{ExoMol} and other databases.   Specifically, \textsc{ExoPhoto} draws its photodissociation cross sections and quantum yields directly from the primary data‐producing projects \textsc{ExoMol}, the \textsc{PhoMol} Team and \textsc{UGAMOP} Database with experimental data from 
 Danish Technical University (DTU, published here for the first time) and the \href{https://www.anr-exact.cnrs.fr/}{\textsc{EXACT} (EXoplanetary Atmospheric Chemistry at High Temperature) database \citep{venotweb}}.

\textsc{ExoPhoto} aims to compile temperature-dependent photodissociation cross section data for molecules relevant to astrophysics, filling gaps identified in previous literature, particularly with regards to temperature-dependence.   By providing a unified format for photodissociation cross section data, this initiative simplifies the process for scientists to reliably compute photodissociation rates and model photon-rich environments, such as the top of exoplanetary atmospheres.

\section{Database Coverage}

The goal of the  \textsc{ExoMol}  project is to include all of the spectroscopic characteristics of molecules that are thought to be significant in hot astrophysical environments. To a certain degree, the required temperature and frequency range completeness depend on what is needed for astronomical and other research such as combustion studies.

Tables \ref{ExoPhotoSummary} list molecules for which the \textsc{ExoPhoto} database currently provides data taken from calculations or measurements, respectively. These datasets are either already freely available in the literature or have been provided directly to us for inclusion in the \textsc{ExoPhoto} database. The calculated data summarised in Table \ref{ExoPhotoSummary} have  been obtained  by the  \textsc{ExoMol}, \textsc{UGAMOP} and \textsc{PhoMol} groups. The experimental data given in Table \ref{ExoPhotoSummary} have been taken from the work of \citet{venot2018vuv,hcch}, see the   \textsc{EXACT} database \citep{venotweb}, and studies performed at the Danish Technical University (DTU) by  Fateev and co-workers. Note that these experiments actually measure UV photoabsorption cross sections rather than explicit photodissociation cross sections;  here, as has been done elsewhere, see \citet{venot2018vuv} for example, we equate these absorption cross sections directly with photodissociation which yields an upper limit and may lead to a slight overestimation of the true photodissociation cross section.
Note that while calculated cross sections are all for a single, specified isotopologue, the experimental data are for isotopologue mixtures in terrestrial natural abundance.

The \textsc{ExoMol} project has developed a specific methodology based on time-independent quantum mechanical calculations using standard bound-state nuclear motion codes to derive photodissociation cross sections \citep{jt840,jt958}. This methodology involves solving the Schr\"{o}dinger equation to determine molecular energy levels, generating potential energy curves (PECs), and identifying transitions involving unbound states \citep{jt969}. For diatomic molecules, the variational nuclear-motion program \textsc{Duo} \citep{jt609} has thus been  employed, while post-processing techniques involves Gaussian smoothing \citep{jt840,jt958, jt976}. We note that \textsc{PhoMol} and \textsc{UGAMOP} use more standard methods which explicitly treat the full set of continuum states; however, thus far \textsc{PhoMol} has not considered photodissociation resulting from predissociation, which can be significant in the important long wavelength (near threshold) region  for some molecules. These methodological differences should be kept in mind when comparing the \textsc{ExoMol} and \textsc{PhoMol} calculations  presented below. 
For the experimentally obtained photodissociation data, the dataset contains some negative values, which likely arise from experimental noise or baseline errors. However, the number of negative values is overall  small and removing them would have the effect of artificially raising the baseline.

\begin{table}
\caption{Summary of calculated and experimental photodissociation cross section data in the \textsc{ExoPhoto} database}
\label{ExoPhotoSummary}
\centering
\begin{tabular}{@{}lrlp{1.4cm}l r c c c l@{}}
\toprule
\textbf{Data Type}
  & \textbf{ID}
  & \textbf{Molecule}
  & \textbf{Isotopologue}
  & \textbf{Dataset}
  & \textbf{$N_{\rm files}$}
  & \makecell[l]{\textbf{Temperature}\\\textbf{Range (K)}}
  & \makecell[l]{\textbf{Wavelength}\\\textbf{Range (nm)}}
  & \makecell[l]{\textbf{Pressure}\\\textbf{Range (bar)}}
  & \textbf{Reference} \\
\midrule
\textbf{Calculated}
  &  1 & AlH       & $^{1}\mathrm{H}^{27}\mathrm{Al}$   & PhoMol      & 209 & 100–10450      & 50–1000.1     & 0         & \cite{qinAlh}                   \\
                    &  2 & HCl     & $^{1}\mathrm{H}^{35}\mathrm{Cl}$   & PTY         &  34 & 0.01–10000     & 100–400       & 0         & \cite{jt865}                   \\
                    &  3 &      &   & PhoMol      &  34 & 0.01–10000     & 50–500        & 0         & \cite{qinHclHF}                \\
                    &  4 & HCl       & $^{1}\mathrm{H}^{37}\mathrm{Cl}$   & PTY         &  34 & 0.01–10000     & 100–400       & 0         & \cite{jt865}                   \\
                    &  5 &        & $^{2}\mathrm{H}^{35}\mathrm{Cl}$   & PTY         &  34 & 0.01–10000     & 100–400       & 0         & \cite{jt865}                   \\
                    &  6 &        & $^{2}\mathrm{H}^{37}\mathrm{Cl}$   & PTY         &  34 & 0.01–10000     & 100–400       & 0         & \cite{jt865}                   \\
                    &  7 & HF        & $^{1}\mathrm{H}^{19}\mathrm{F}$    & PTY         &  34 & 0.01–10000     & 90–400.1      & 0         & \cite{jt865}                   \\
                    &  8 &        &    & PhoMol      &  34 & 0.01–10000     & 50–500        & 0         & \cite{qinHclHF}                \\
                    &  9 &         & $^{2}\mathrm{H}^{19}\mathrm{F}$    & PTY         &  34 & 0.01–10000     & 90–400.1      & 0         & \cite{jt865}                   \\
                    & 10 & MgH       & $^{24}\mathrm{Mg}^{1}\mathrm{H}$   & UGAMOP      &  10 & 1000–10000     & 170–454       & 0         & \cite{ugamgh1,ugamgh2,ugamgh3} \\
                    & 11 & OH        & $^{16}\mathrm{O}^{1}\mathrm{H}$    & MYTHOS      &  80 & 0.01–7900      & 82.8–2000     & 0         & \cite{jt982}           \\
                    & 12 & NaO       & $^{23}\mathrm{Na}^{16}\mathrm{O}$  & PhoMol      &  34 & 0.01–10000     & 50–500        & 0         & \cite{qinNaO}                  \\
                    & 13 & MgO       & $^{24}\mathrm{Mg}^{16}\mathrm{O}$  & PhoMol      &  35 & 0.01–10000     & 50–1000.1     & 0         & \cite{qinMgO}                  \\
                    & 14 & O$_2$     & $^{16}\mathrm{O}^{16}\mathrm{O}$   & PhoMol      & 101 & 0–10000        & 50–500        & 0         & \cite{qinO2}                   \\
                    & 15 & AlCl      & $^{27}\mathrm{Al}^{35}\mathrm{Cl}$ & PhoMol      & 201 & 0.01–10000     & 50–499        & 0         & \cite{qinAlCl}                 \\
                    & 16 & AlF       & $^{27}\mathrm{Al}^{19}\mathrm{F}$  & PhoMol      & 201 & 0.01–10000     & 50–1000.1     & 0         & \cite{qinAlF}                  \\
                    & 17 & CS        & $^{12}\mathrm{C}^{32}\mathrm{S}$   & UGAMOP      &  10 & 1000–10000     & 50–5000       & 0         & \cite{ugaCS}                   \\
                    & 18 & HeH$^+$   & $^{1}\mathrm{H}\,{}^{4}\mathrm{He}$ & UGAMOP     &   9 & 500–12000      & 10–112.7      & 0         & \cite{ugaHeH}                  \\

\textbf{Experimental}
                    & 19 & CO        & -                             & DTU         &   2 & 305–1630       & 117.04–228.8  & 1–1.0658  & –                             \\
                    & 20 & CO$_2$    & -                            & DTU         &   5 & 305–1630       & 108.79–323.79 & 1–1.0647  & –                             \\
                    & 21 & CO$_2$    &    -                               & EXACT       &   9 & 150–800        & 114–230       & 1         & \cite{venot2018vuv}           \\
                    & 22 & H$_2$O    &-                            & DTU         &   5 & 423.15–1773.15 & 108.81–236.81 & 1         & –                             \\
                    & 23 & SO$_2$    & -                           & DTU         &   1 & 423.15         & 110.35–230.36 & 1         & –                             \\
                    & 24 & C$_2$H$_2$& -                        & photo-FPBV  &   6 & 296–773        & 116–228       & 1         & \cite{hcch}                   \\
                    & 25 & C$_2$H$_4$& -                   & DTU         &   1 & 562.15         & 113.29–201.25 & 1         & –                             \\
                    & 26 & H$_2$CO   & -                       & DTU         &   4 & 303.15–573.15  & 110–230.014   & 1         & –                             \\
                    & 27 & NH$_3$    & -                           & DTU         &   2 & 289–295.55     & 113.28–201.25 & 1         & –                             \\
\bottomrule
\end{tabular}
\end{table}

\subsection{Diatomics}
\subsubsection{\upshape AlCl (PhoMol) \citep{qinAlCl}}

\begin{figure}
    \centering
   \includegraphics[width=1\textwidth]{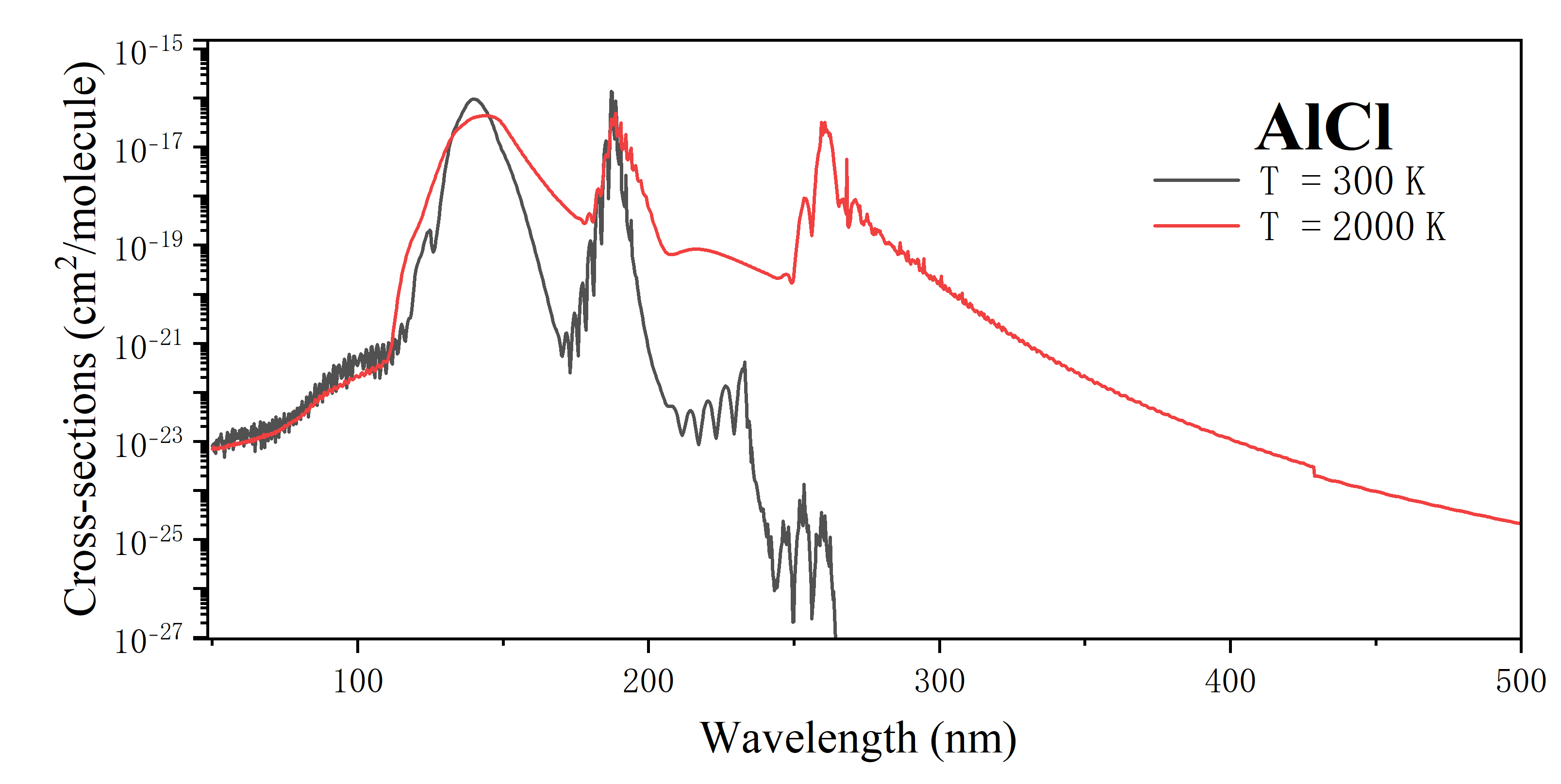} 
    \caption{AlCl  from \textsc{PhoMol} spectrum overview \citep{qinAlCl}.} 
    \label{fig:AlCl}
\end{figure}

Photodissociation cross sections for the single isotopologue \(^{27}\mathrm{Al}^{35}\mathrm{Cl}\) were computed with \textsc{PhoMol} over a wavelength range of 50–499\,nm and for temperatures from 0.01\,K to 10\,000\,K; example results at two temperatures are shown in Fig.~\ref{fig:AlCl}. The required potential energy curves and transition dipole moments for the ground X\(^1\Sigma^+\) state and six excited singlet states were obtained via HF + CASSCF/aug-cc-pV6Z followed by icMRCI+Q/aug-cc-pV6Z. State‐resolved cross sections from all 38\,958 rovibrational levels were then determined by numerically solving the radial Schr\"{o}dinger equation across photon wavelengths from 50\,nm to the dissociation threshold.


\subsubsection{\upshape AlF (PhoMol) \citep{qinAlF}}

\begin{figure}
    \centering
    \includegraphics[width=0.9\textwidth]{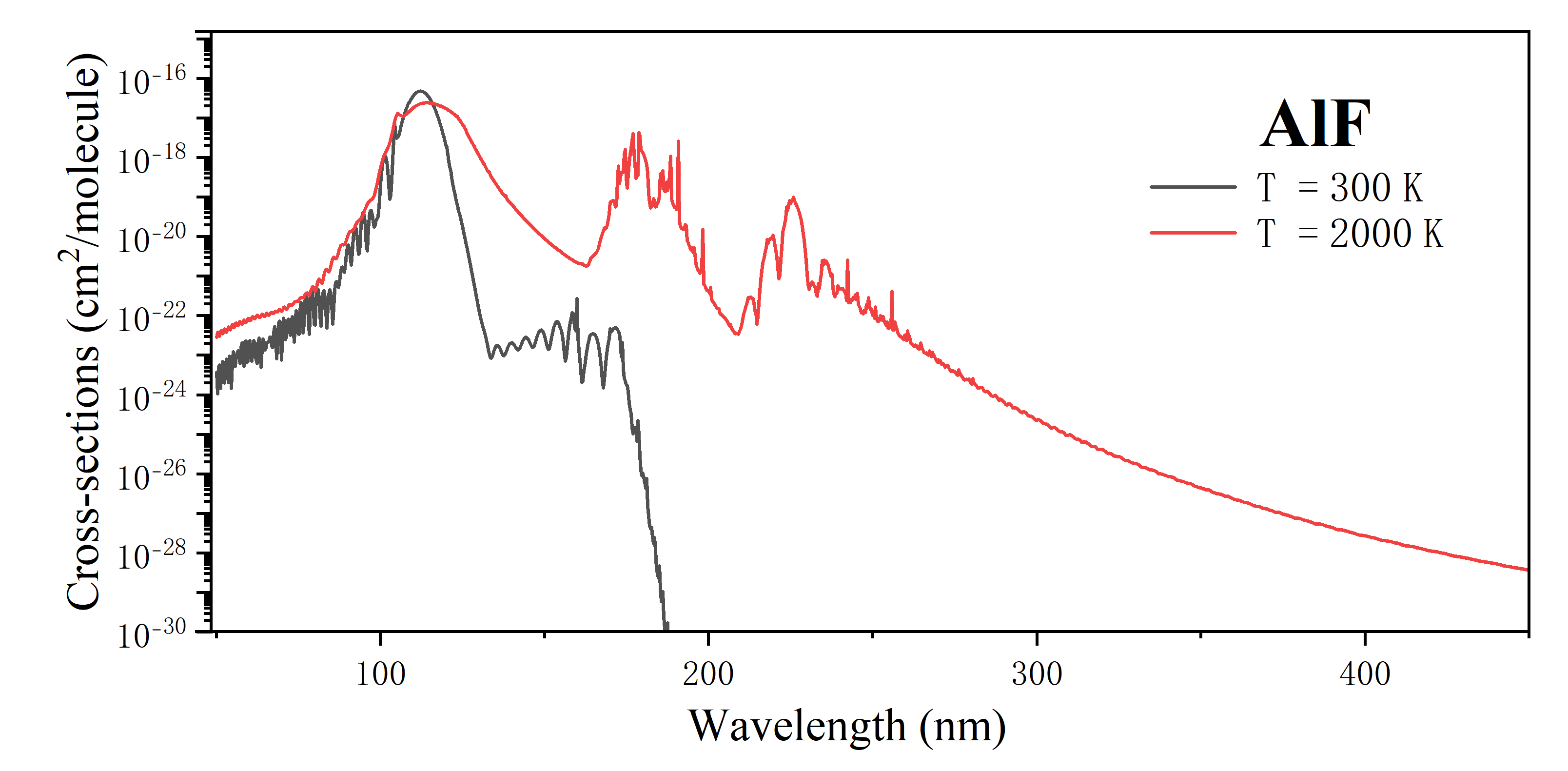} 
    \caption{AlF  spectrum overview from \textsc{PhoMol} \citep{qinAlF}.} 
    \label{fig:AlF}
\end{figure}

For aluminum monofluoride (\(^{27}\mathrm{Al}^{19}\mathrm{F}\)), photodissociation cross sections were computed with \textsc{PhoMol} over the wavelength range 50.0–1000.1\,nm for temperatures from 0.01\,K to 10\,000\,K; example results at two temperatures appear in Fig.~\ref{fig:AlF}. Potential energy curves and transition dipole moments for the ground X\(^1\Sigma^+\) state and seven excited singlet states were calculated via icMRCI+Q with the aug-cc-pCV5Z-DK basis set. State‐resolved cross sections from 36\,349 rovibrational levels of the X\(^1\Sigma^+\) state were then obtained by numerically solving the continuum wavefunctions. These data have been applied to carbon-star envelope models, where AlF abundance is governed by photodissociation.

\subsubsection{\upshape AlH (PhoMol) \citep{qinAlh}}

\begin{figure}
    \centering
    \includegraphics[width=0.9\textwidth]{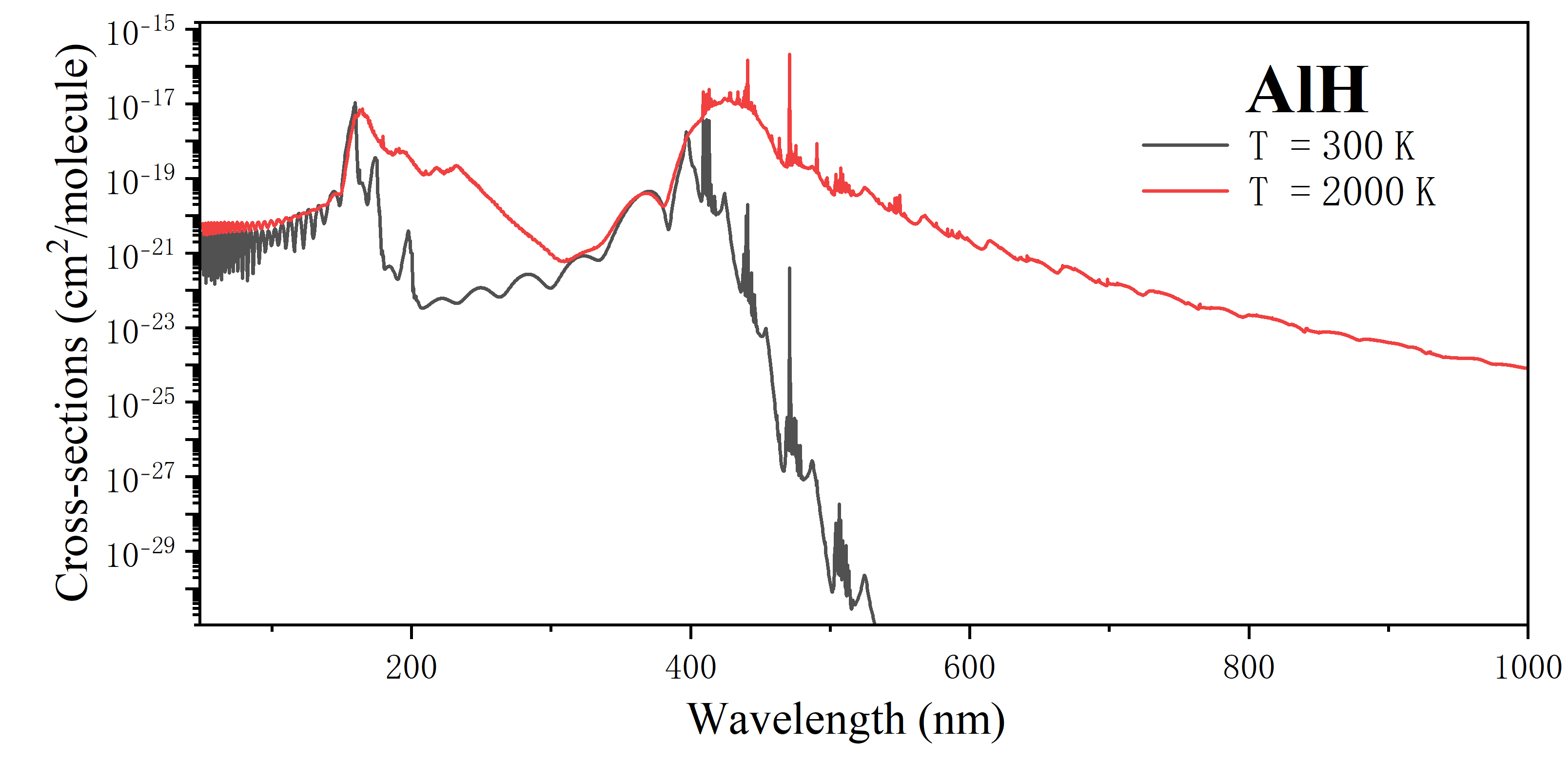} 
    \caption{AlH  spectrum overview \citep{qinAlh}.} 
    \label{fig:AlH}
\end{figure}

The aluminum hydride (\(^{27}\mathrm{Al}^{1}\mathrm{H}\)) molecule's photodissociation cross sections were calculated  by \textsc{PhoMol}, see Fig.~\ref{fig:AlH}. Wavelengths range from 50 nm to 1000.1 nm, and calculations were performed for temperatures from 100 K to 10~450 K. We note these cross sections neglect the effects of predissociation which studies have shown to be important for  AlH \citep{jt874,jt922} and their inclusion is likely to significantly increase the photodissociation cross section at long wavelengths.

\subsubsection{\upshape CO (DTU)}

\begin{figure}
    \centering
    \includegraphics[width=0.9\textwidth]{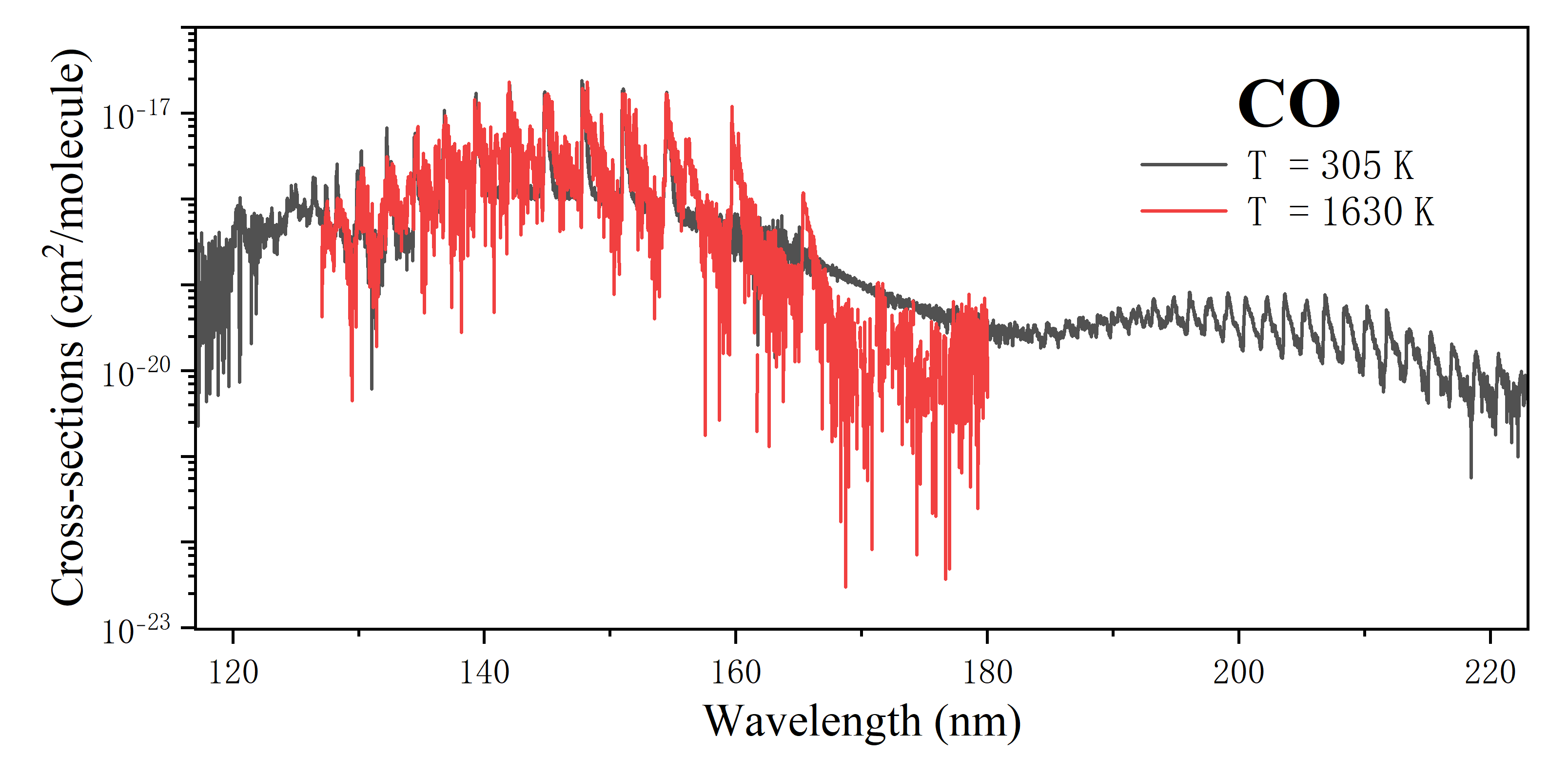} 
    \caption{CO  spectrum overview \citep{Fateev2021}} 
    \label{fig:CO}
\end{figure}

Photodissociation cross sections for carbon monoxide (CO) were recorded in the far-UV region at the Danish Technical University (DTU) by  Fateev and co-workers. The data span a wavelength range from 117.04~nm to 228.8~nm and are provided for two temperatures, 1630~K and 305~K, for pressures of 1 to 1.0658~bar, see Fig.~\ref{fig:CO}. 

\subsubsection{\upshape CS (UGAMOP) \citep{ugaCS}}

\begin{figure}
    \centering
    \includegraphics[width=0.9\textwidth]{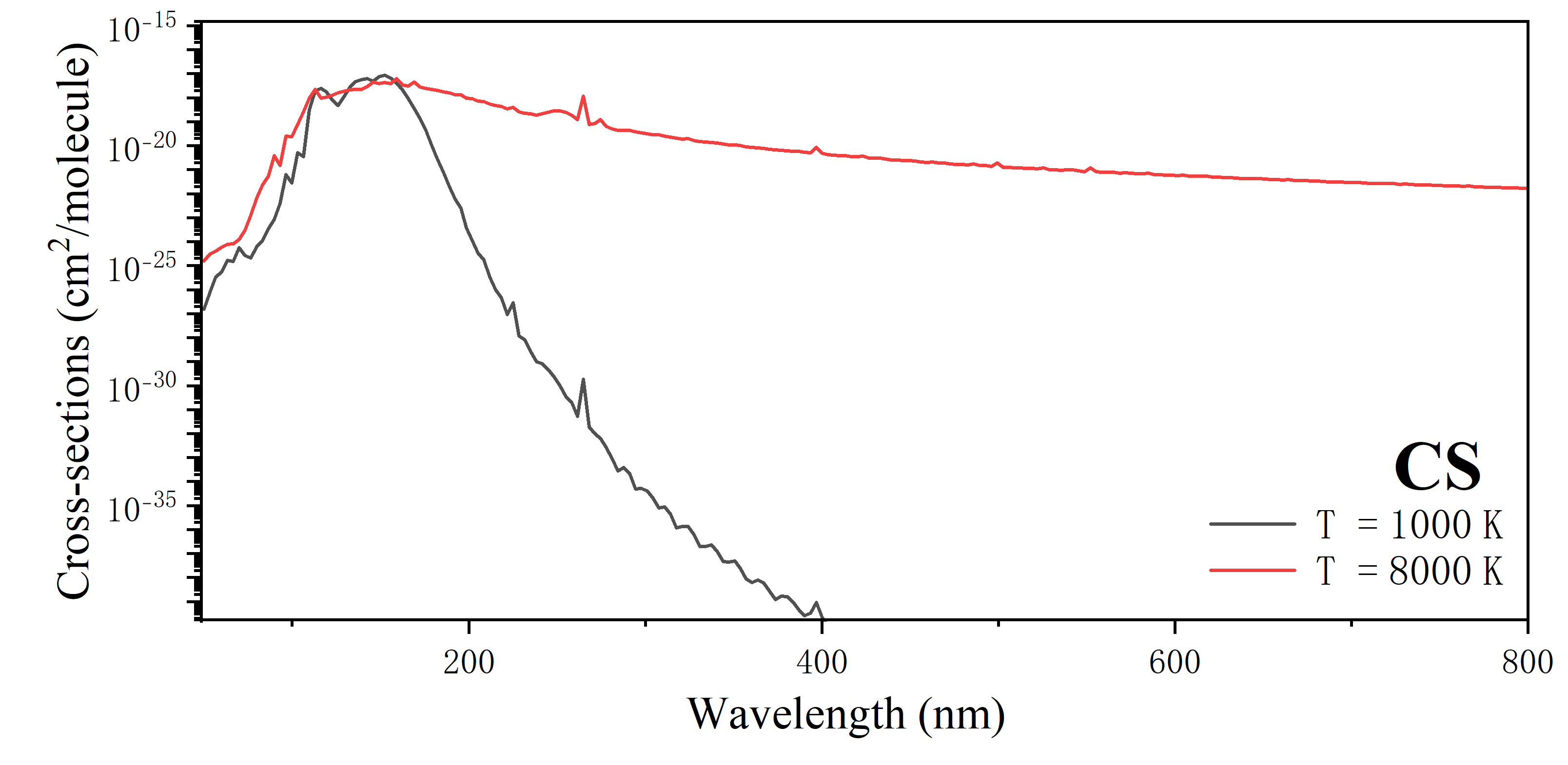} 
    \caption{CS  spectrum overview \citep{ugaCS}} 
    \label{fig:CS}
\end{figure}

The photodissociation cross sections of carbon monosulfide (\(^{12}\mathrm{C}^{32}\mathrm{S}\)) were calculated using quantum-mechanical techniques by \textsc{UGAMOP}, see Fig.~\ref{fig:CS}. Transitions were modeled from the \( X\,^1\Sigma^+ \) electronic ground state to six low-lying excited states (\( A\,^1\Pi \), \( A'\,^1\Sigma^+ \), \( 2\,^1\Pi \), \( 3\,^1\Pi \), \( B\,^1\Sigma^+ \), and \( 4\,^1\Pi \)). Cross sections cover a wavelength range of 50 nm to 5000 nm and were evaluated for temperatures between 1000~K and 10,000~K. These results provide essential insights for astrophysical applications, including dense interstellar clouds, planetary nebulae, and photodissociation regions.

\subsubsection{\upshape HCl and HF (ExoMol and PhoMol) \citep{jt865,qinHclHF}}

The \textsc{ExoMol} project computed photodissociation cross sections for HCl and HF using high-accuracy potential energy curves (PECs) and dipole moment curves (DMCs) from advanced quantum mechanical methods. Calculations employed the Duo variational nuclear motion program \citep{jt609} to solve the Schr\"{o}dinger equation for both bound and unbound states. The Gaussian-smoothed cross sections were validated against available experimental data. These results can be used to model UV-driven processes in diverse astrophysical and planetary environments, particularly in hot and UV-rich regions.

For HCl, the \textsc{ExoMol} photodissociation database (PTY) provides cross sections for four isotopologues: \textsuperscript{1}H\textsuperscript{35}Cl, \textsuperscript{1}H\textsuperscript{37}Cl, \textsuperscript{2}H\textsuperscript{35}Cl, and \textsuperscript{2}H\textsuperscript{37}Cl. These cross sections span a wavelength range of 100~nm to 400~nm and are evaluated across 34 temperatures, ranging from 0.01~K to 10~0000,000~K. The temperature values correspond to excitation conditions of the gas, assuming local thermal equilibrium (LTE) and the Boltzmann distribution for the population of energy states. The total cross sections were obtained by summing contributions from all final electronic states. 

\textsc{PhoMol} also includes cross sections for \textsuperscript{1}H\textsuperscript{35}Cl, calculated over the wavelength and temperature ranges 50~nm to 500~nm and 0.01~K to 10~00,000~K. Temperature-dependent cross sections are evaluated across 34 temperatures.

Similarly, HF photodissociation cross sections in the \textsc{ExoMol} PTY database were computed using the same methodology. The database provides data for two isotopologues: \textsuperscript{1}H\textsuperscript{19}F and \textsuperscript{2}H\textsuperscript{19}F. These calculations account for transitions involving both bound and unbound states, with cross sections covering a wavelength range of 90~nm to 400.1~nm and evaluated over the same temperature grid (0.01~K to 10 000~K).

\textsc{PhoMol} also includes \textsuperscript{1}H\textsuperscript{19}F data, with the 90--400.1 nm wavelength and 0.01--10000 K temperature ranges as PTY. The database includes seven electronic transitions with state-resolved cross sections derived from all rovibrational levels of the \(X^1\Sigma^+\) state.

\begin{figure}
    \centering
    \includegraphics[width=0.8\textwidth]{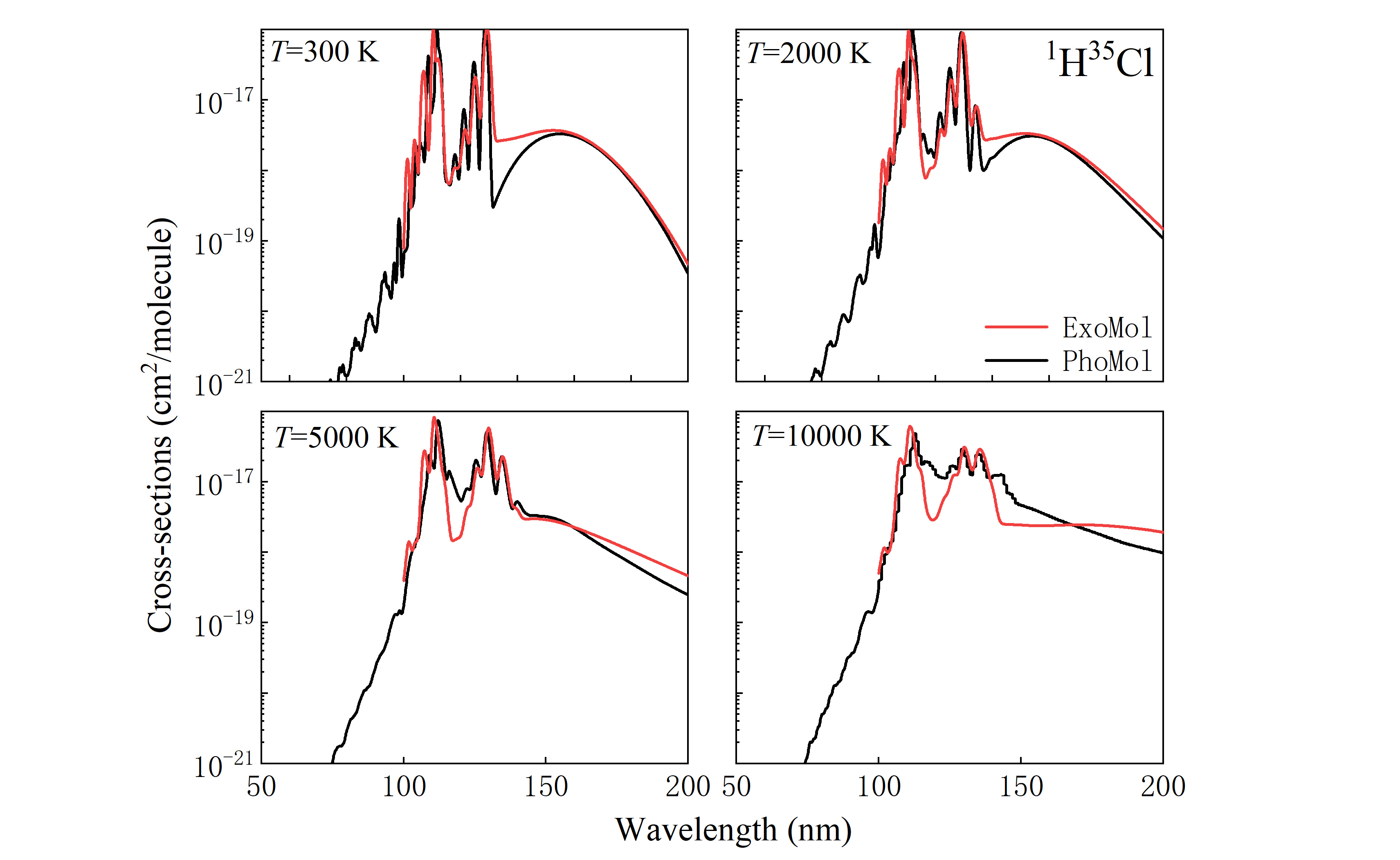} 
    \caption{\textsuperscript{1}H\textsuperscript{35}Cl  spectrum overview for \textsc{ExoMol} and \textsc{PhoMol} \citep{jt865,qinHclHF}} 
    \label{fig:1H35Cl}
\end{figure}

\begin{figure}
    \centering
    \includegraphics[width=0.8\textwidth]{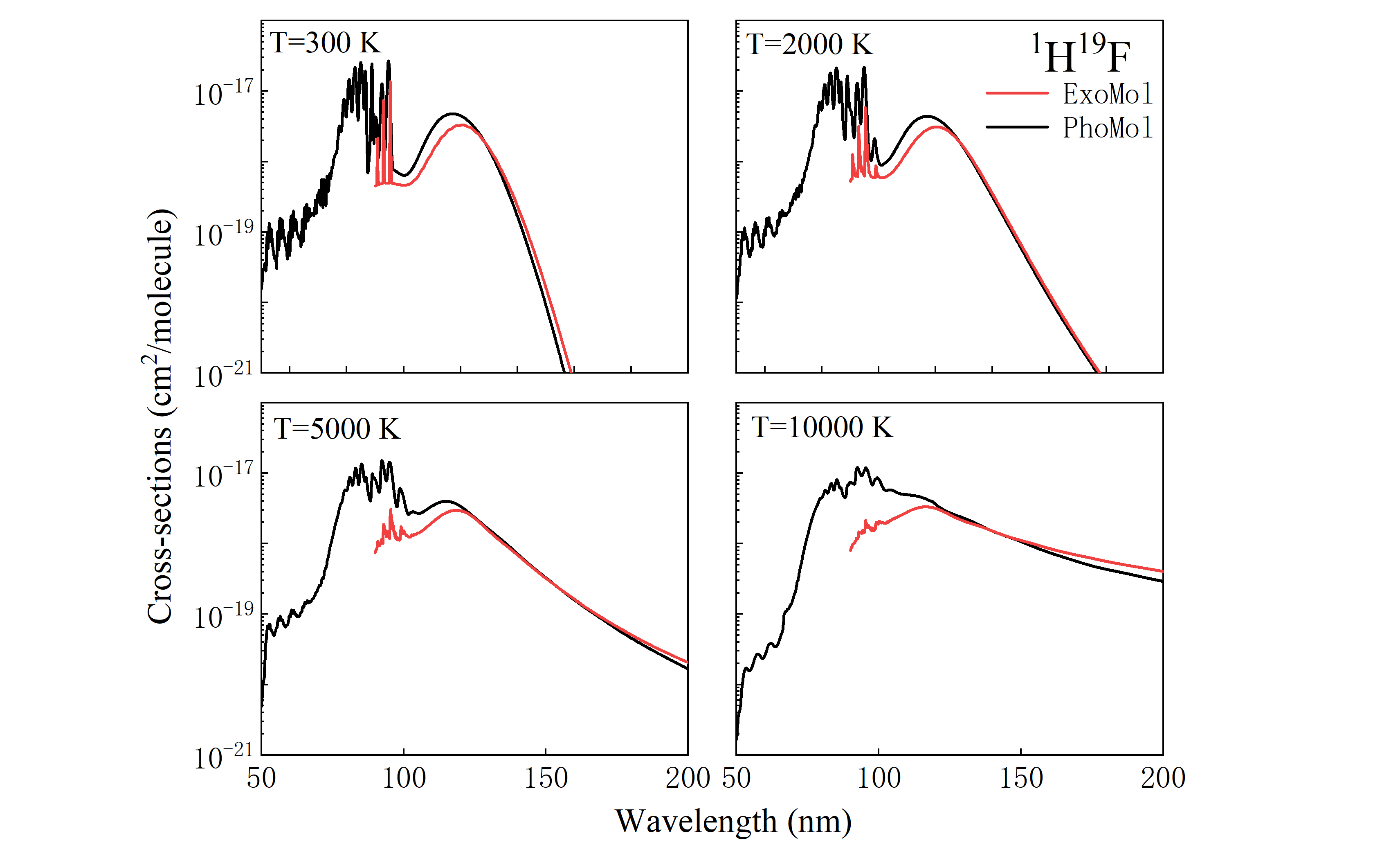} 
    \caption{\textsuperscript{1}H\textsuperscript{19}F  spectrum overview for \textsc{ExoMol} and \textsc{PhoMol} \citep{jt865,qinHclHF}} 
    \label{fig:1H19F}
\end{figure}

\begin{figure}
    \centering
    \includegraphics[width=0.8\textwidth]{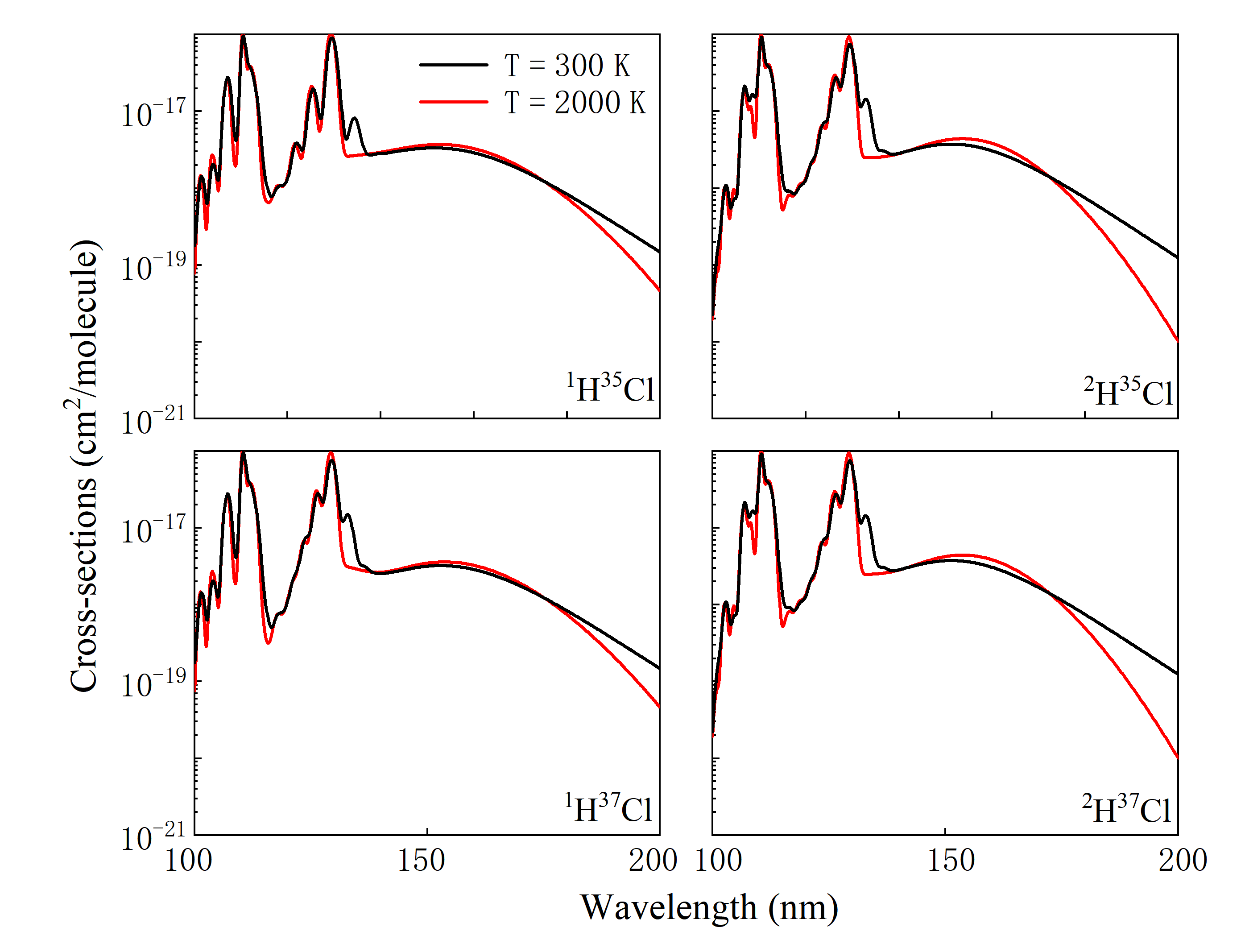} 
    \caption{HCl  spectrum overview for \textsc{ExoMol} \citep{jt865}.} 
    \label{fig:HCl}
\end{figure}

\begin{figure}
    \centering
    \includegraphics[width=0.6\textwidth]{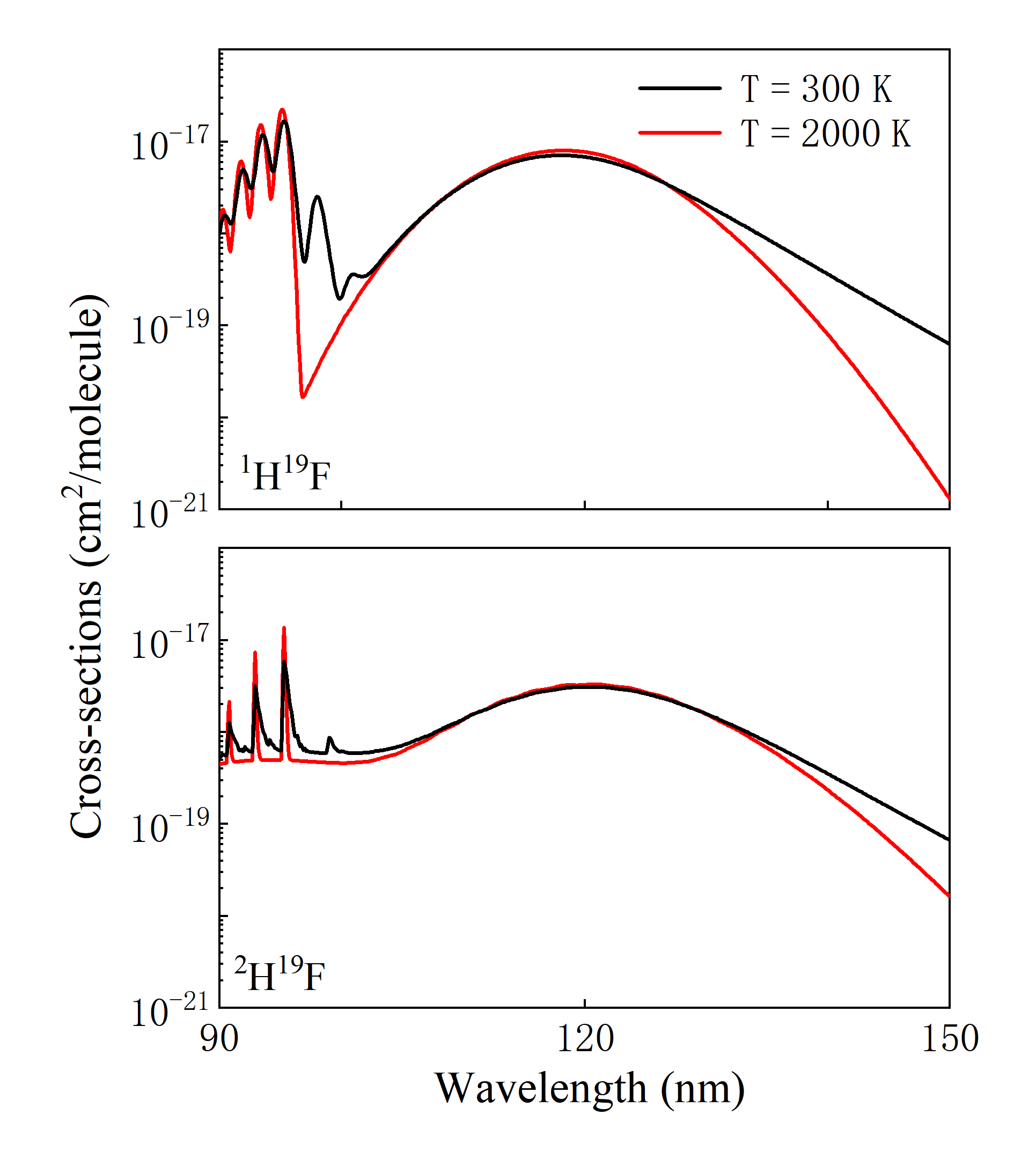} 
    \caption{HF  spectrum overview from \textsc{ExoMol} \citep{jt865}} 
    \label{fig:HF}
\end{figure}

The \textsc{ExoMol} and \textsc{PhoMol} results for the two species are in very good agreement, see Figs. \ref{fig:1H35Cl} and \ref{fig:1H19F}. All the HCl isotopologue  spectrum overviews from \textsc{ExoMol} are shown in Fig. \ref{fig:HCl}. A spectrum overview for  \textsuperscript{1}H\textsuperscript{19}F and \textsuperscript{2}H\textsuperscript{19}F isotopologues from ExoMol   is shown in Fig. \ref{fig:HF}.

Both databases provide critical cross section data for studying photodissociation processes in interstellar and planetary environments.
 We therefore recommend the \textsc{ExoMol} datasets as these provide cross sections for all
the stable isotopologues but note that those interested in short wavelengths it will need to use the \textsc{PhoMol} results.

\subsubsection{ \upshape HeH\textsuperscript{+} (UGAMOP) \citep{ugaHeH}}

\begin{figure}
    \centering
    \includegraphics[width=0.9\textwidth]{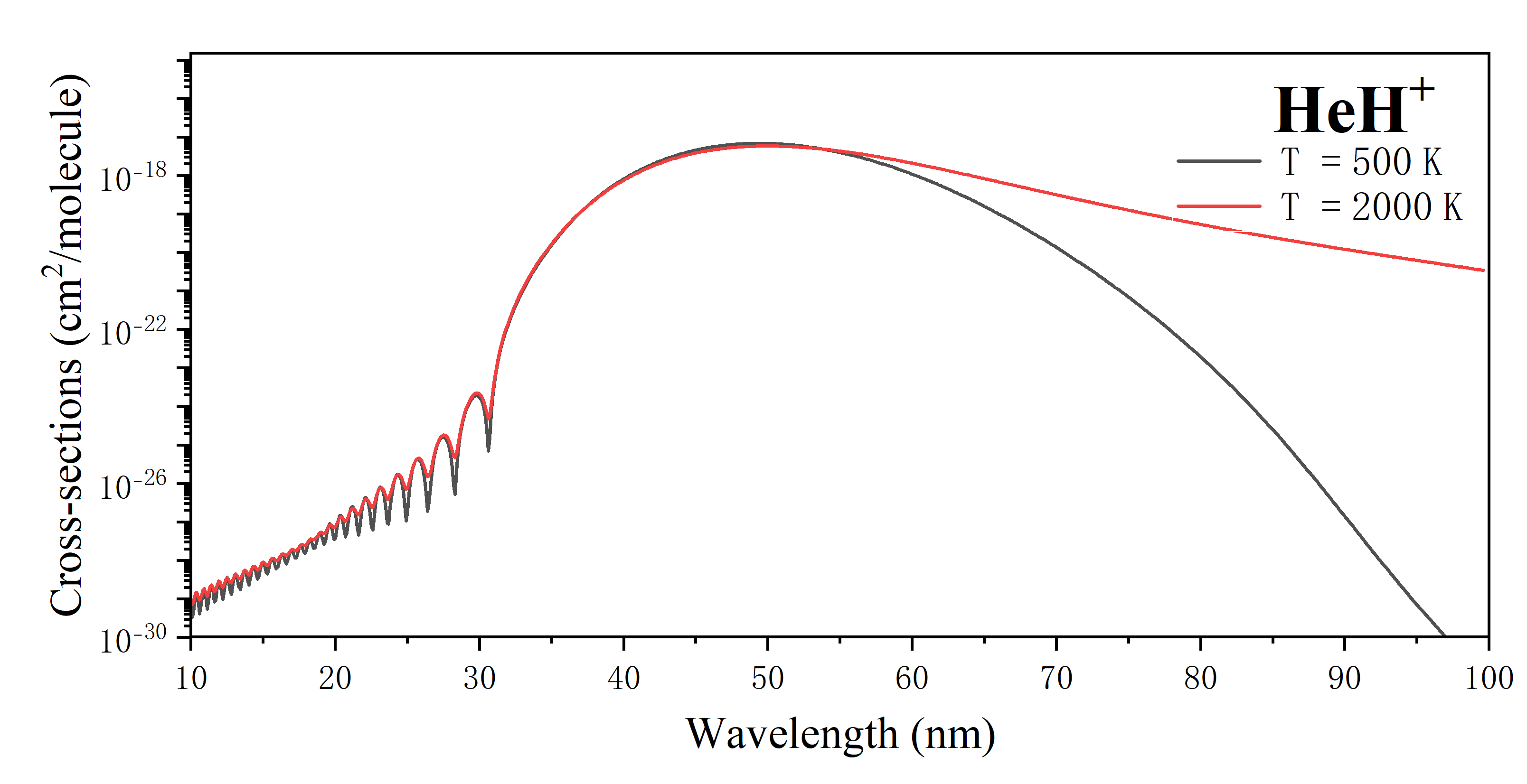} 
    \caption{HeH\textsuperscript{+}  spectrum overview \citep{ugaHeH}} 
    \label{fig:HeH+}
\end{figure}

Accurate photodissociation cross sections for the helium hydride ion (\textsuperscript{4}He\textsuperscript{1}H\textsuperscript{+}) were calculated  by \textsc{UGAMOP}. The calculations covered all vibrational levels (\( v'' = 0 \) to 11). They are illustrated in Fig.~\ref{fig:HeH+}.


\subsubsection{\upshape MgH (UGAMOP) \citep{ugamgh1,ugamgh2,ugamgh3}}

\begin{figure}
    \centering
    \includegraphics[width=0.9\textwidth]{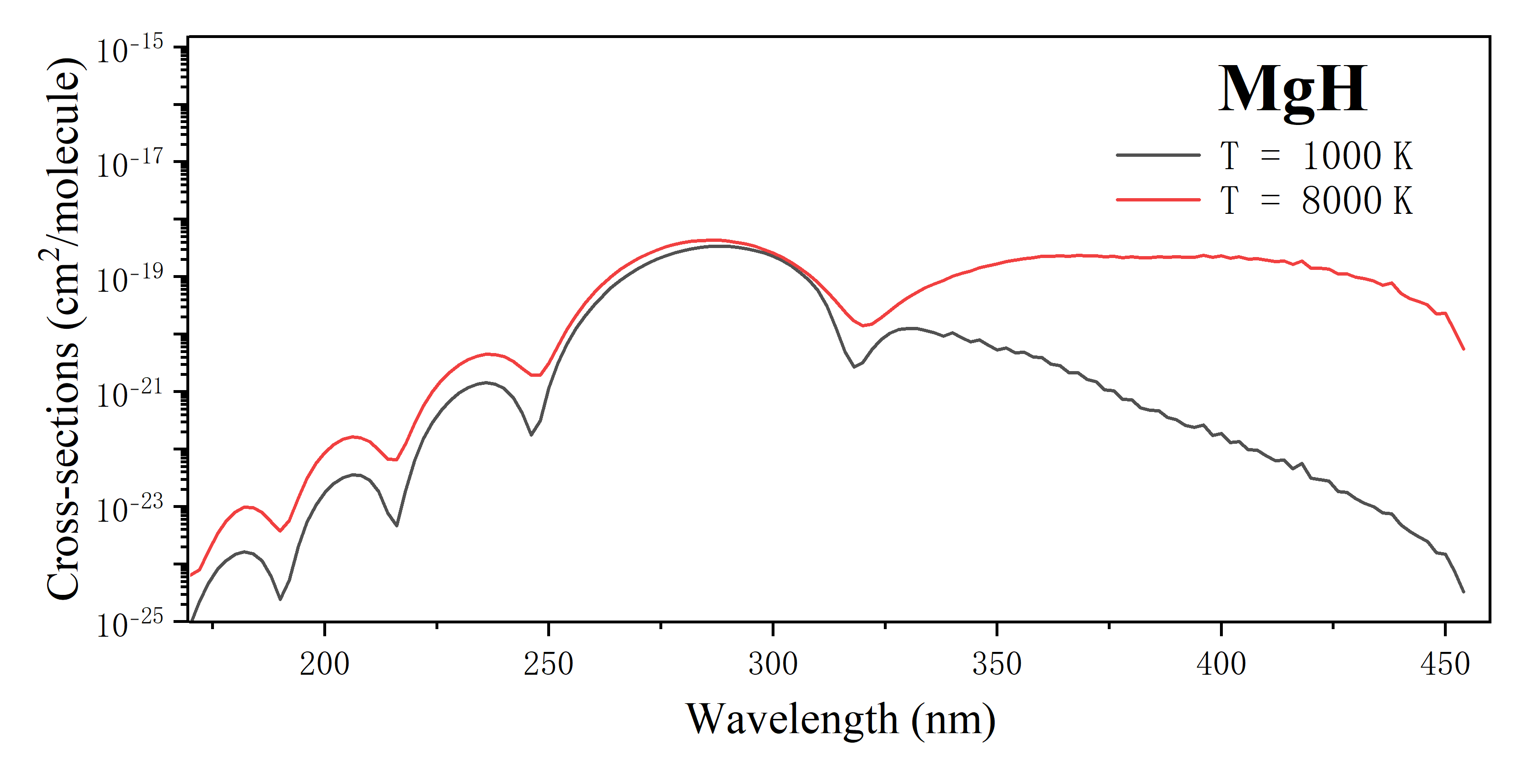} 
    \caption{MgH  spectrum overview \citep{ugamgh1,ugamgh2,ugamgh3}} 
    \label{fig:MgH}
\end{figure}

Photodissociation cross sections for \(^{24}\mathrm{Mg}^{1}\mathrm{H}\) were calculated by \textsc{UGAMOP} using high-accuracy PECs and DMCs derived from a combination of \textit{ab initio} and experimental methods. The cross sections primarily focus on the \(B'^2\Sigma^+ \leftarrow X^2\Sigma^+\) and \(A^2\Pi \leftarrow X^2\Sigma^+\) electronic transitions, covering rovibrational levels of the ground state. Cross sections span the wavelength range from 170~nm to 454~nm and are provided for temperatures between 1\,000~K and 100\,000\,000~K. UGAMOP cross sections of MgH for $T=300$ and $2000$~K are illustrated in Fig.~\ref{fig:MgH}. 

\subsubsection{\upshape MgO (PhoMol) \citep{qinMgO}}

\begin{figure}
    \centering
    \includegraphics[width=0.9\textwidth]{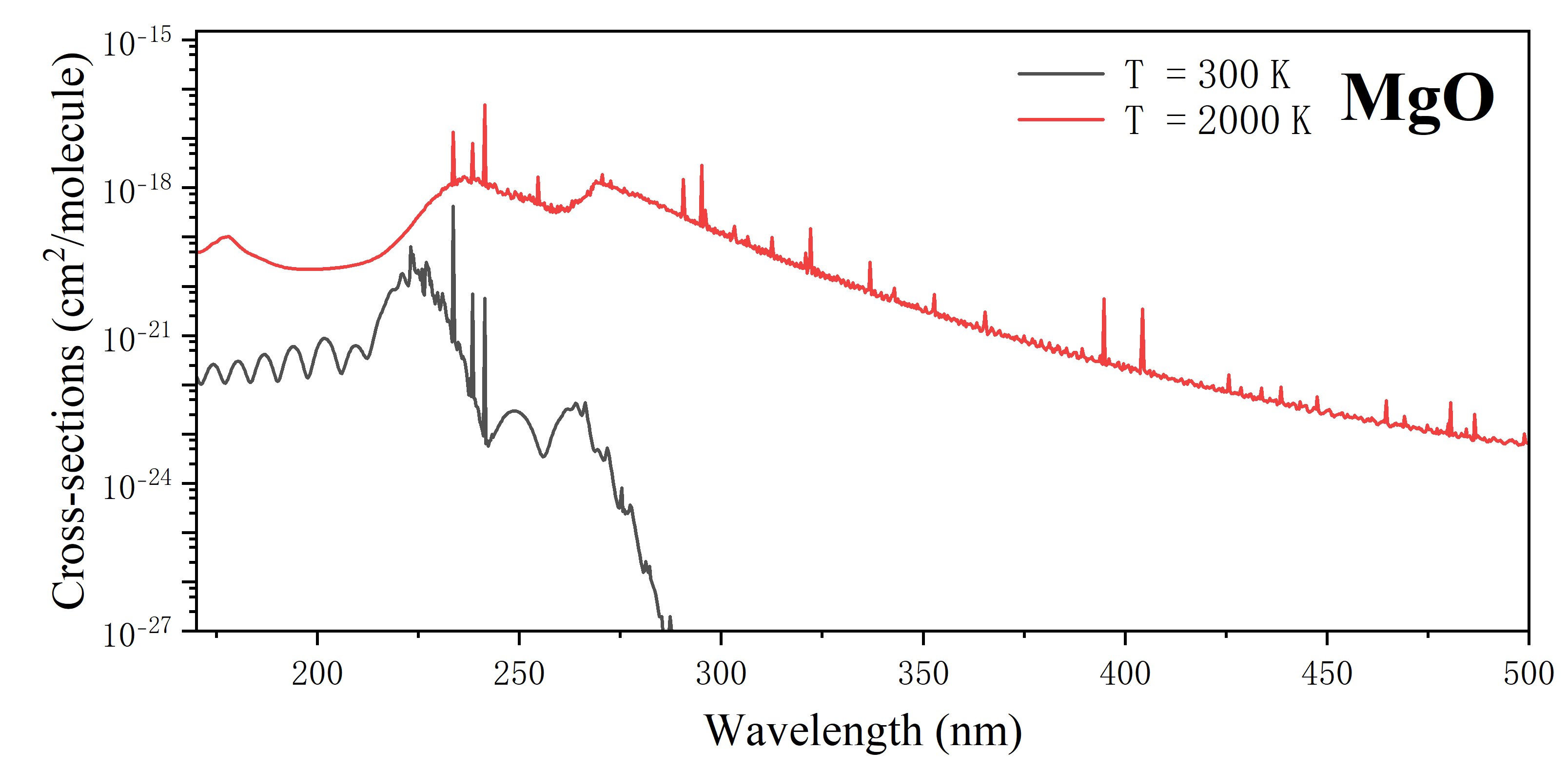} 
    \caption{MgO photodissociation cross sections \citep{qinMgO}} 
    \label{fig:MgO}
\end{figure}

Magnesium oxide (\(^{24}\mathrm{Mg}^{16}\mathrm{O}\)) photodissociation cross sections were computed by \textsc{PhoMol} for transitions from the \(X^1\Sigma^+\) state.   \citet{qinMgO} note that the photodissociation cross sections from the \(a^3\Pi\) state are  questionable due to an incorrect treatment of its partition function; these cross sections are therefore omitted here. The wavelength range covers 50–500\,nm, and cross sections were evaluated for temperatures between 0.01\,K and 10\,000\,K; example results at two temperatures are shown in Fig.~\ref{fig:MgO}. As with other \textsc{PhoMol} datasets, this study considers only direct continuum dissociation and does not include any predissociation contributions, which may be significant for MgO. These results are important for modeling magnesium chemistry in planetary exospheres and the envelopes of evolved stars.

\subsubsection{\upshape NaO (PhoMol) \citep{qinNaO}}

\begin{figure}
    \centering
    \includegraphics[width=0.9\textwidth]{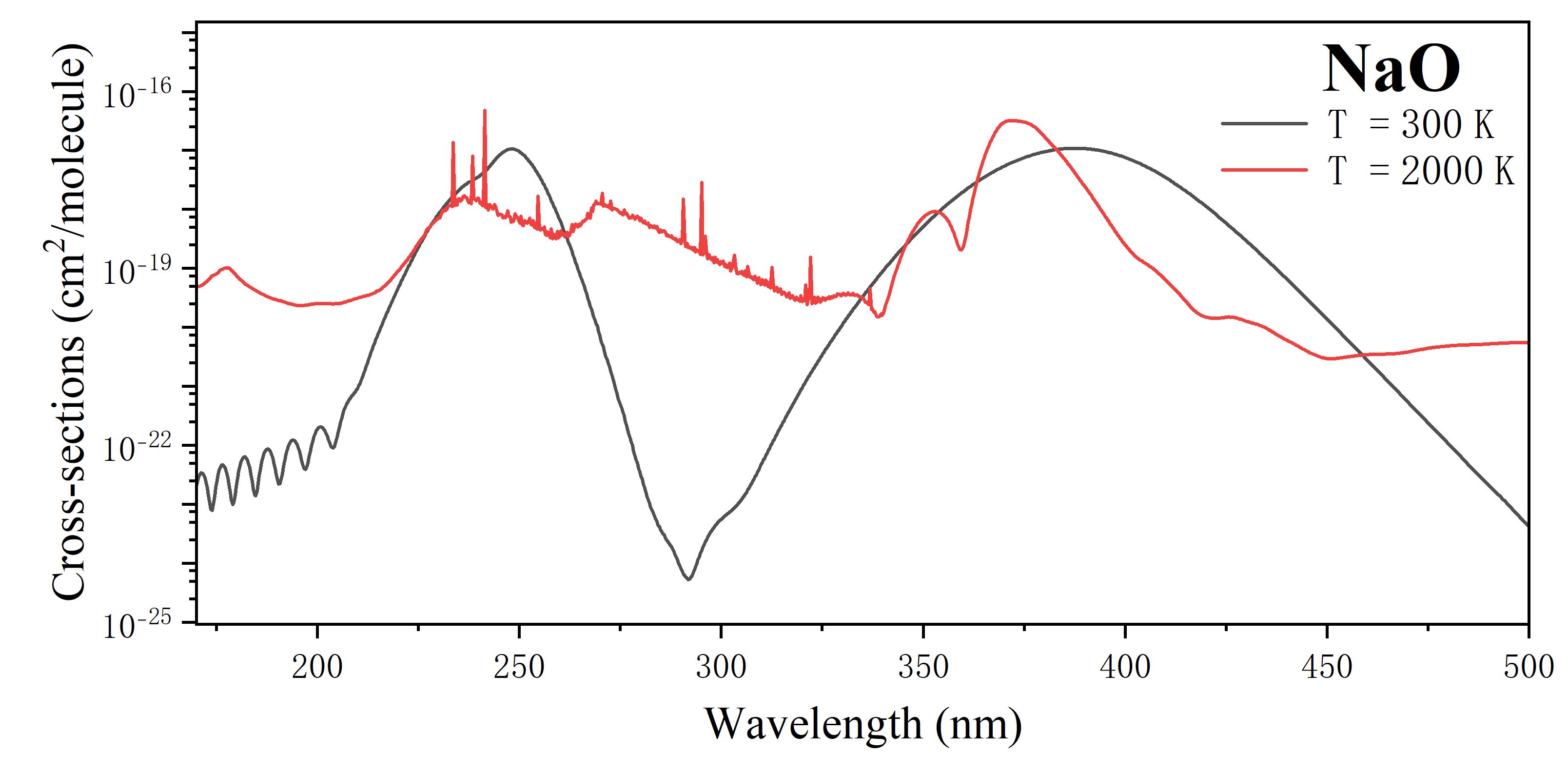} 
    \caption{NaO  spectrum overview \citep{qinNaO}} 
    \label{fig:NaO}
\end{figure}

The photodissociation cross sections of sodium oxide (\(^{23}\mathrm{Na}^{16}\mathrm{O}\)) were computed with \textsc{PhoMol} for transitions from the X$^2\Sigma^+$ and A$^2\Sigma^+$ states over the wavelength range 50–500\,nm and temperatures 0.01–10\,000\,K; example results at two temperatures are shown in Fig.~\ref{fig:NaO}. PECs and DMCs for the seven lowest electronic states were obtained via CASSCF followed by valence icMRCI+Q with the aug-cc-pCV6Z basis set. State‐resolved cross sections were then derived by numerically solving the nuclear Schr\"{o}dinger equation across all bound rovibrational levels. However, as with other \textsc{PhoMol} datasets, only direct continuum dissociation is treated and predissociation contributions are omitted. These data are essential for modeling sodium photochemistry in planetary exospheres and cool stellar atmospheres.

\subsubsection{\upshape OH (ExoMol) \citep{jt982}}

\begin{figure}
    \centering
    \includegraphics[width=0.9\textwidth]{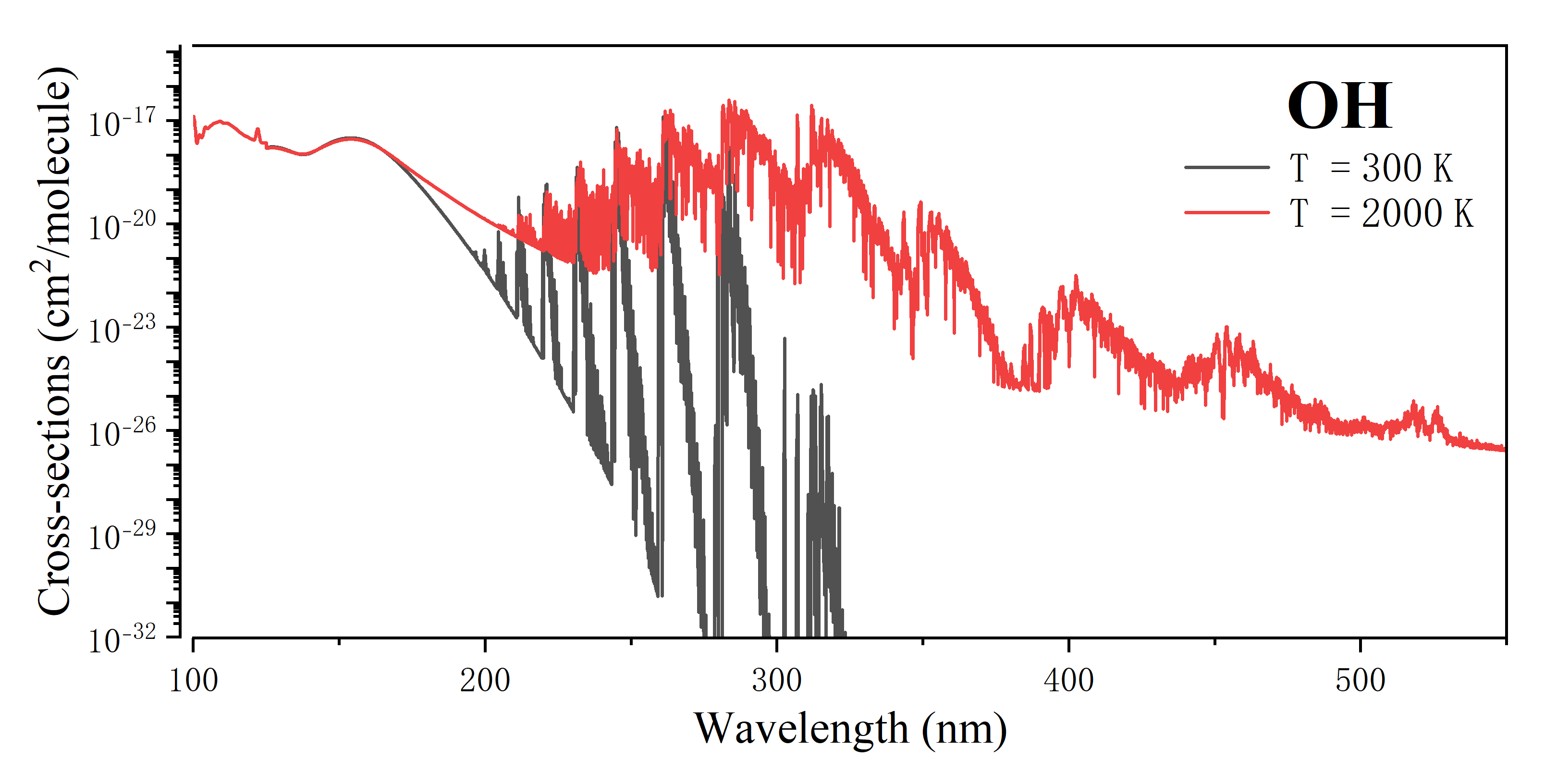} 
    \caption{OH  spectrum overview \citep{jt982}} 
    \label{fig:OH}
\end{figure}

The photodissociation cross sections for the hydroxyl radical (OH) by \textsc{ExoMol} adopted the main features of the spectroscopic model used to compute the \textsc{MYTHOS} line list  \citep{jt969}; this model used calculated  using high-accuracy multi-reference configuration interaction (MRCI) PECs and couplings, adjusted to available experimental data \cite{jt869} where available, and transition DMCs.  The model incorporated key electronic states, including \(X\,^2\Pi\), \(A\,^2\Sigma^+\), and repulsive states such as \(1\,^2\Sigma^-\), \(1\,^4\Sigma^-\), and \(1\,^4\Pi\) for \textsuperscript{16}O\textsuperscript{1}H; cross sections were computed over a wavelength range from 82.8~nm to 2000~nm and evaluated for 81 temperatures ranging from 0~K to 8,000~K. Gaussian smoothing was applied to ensure continuous profiles, and the results were tested against cross sections in the Leiden database. The OH cross sections are illustrated in Fig.~\ref{fig:OH}. These cross sections provide essential data for studying UV-driven processes in atmospheric and astrophysical environments. 

We are working on extensions to the \textsc{ExoPhoto} database to allow for the inclusion the photodissociation cross sections needed for studies of non-local
thermodynamic equilibrium environments, which are important for OH. At present these data are available up request to the corresponding author.

\subsubsection{\upshape O$_2$ (PhoMol) \citep{qinO2}}

\begin{figure}
    \centering
    \includegraphics[width=0.9\textwidth]{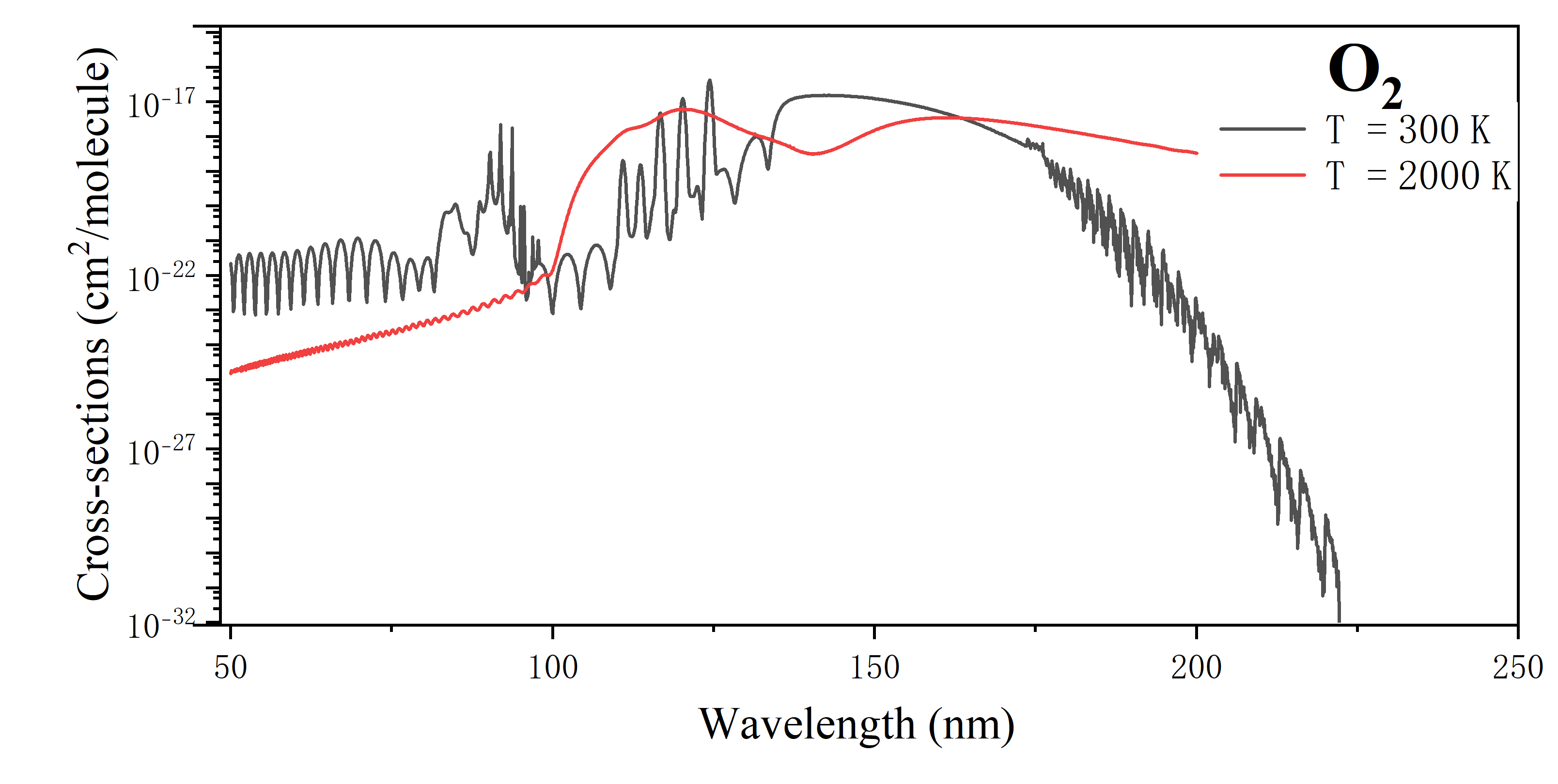} 
    \caption{O$_2$  spectrum overview \citep{qinO2}} 
    \label{fig:O2}
\end{figure}

The photodissociation cross sections for dioxygen (\(^{16}\mathrm{O_2}\)) were computed with \textsc{PhoMol} for photon wavelengths from 50 \AA\ to the relevant dissociation thresholds (approximately 50-500 nm) and for gas temperatures 0–10 000 K; example results at two temperatures are shown in Fig.~\ref{fig:O2}. PECs and DMCs for the four key photodissociation channels (X\(^3\Sigma_g^-\)→B\(^3\Sigma_u^-\), X\(^3\Sigma_g^-\) $\to$ E\(^3\Sigma_u^-\), a\(^1\Delta_g\)→\(^1\Pi_u\), b\(^1\Sigma_g^+\)→\(^1\Pi_u\)) were obtained at the icMRCI+Q/aug-cc-pwCV5Z-DK level, and state‐resolved cross sections were derived by numerically solving the nuclear Schr\"{o}dinger equation followed by Boltzmann averaging over 0–10 000 K.

\subsection{Triatomics}

\subsubsection{\upshape CO\textsubscript{2} (DTU and EXACT) \citep{Fateev2021,venot2018vuv}}

The \textsc{ExoPhoto} database provides two comprehensive datasets offering temperature- and pressure-dependent absorption properties for CO$_2$.

\begin{figure}
    \centering
    \includegraphics[width=0.7\textwidth]{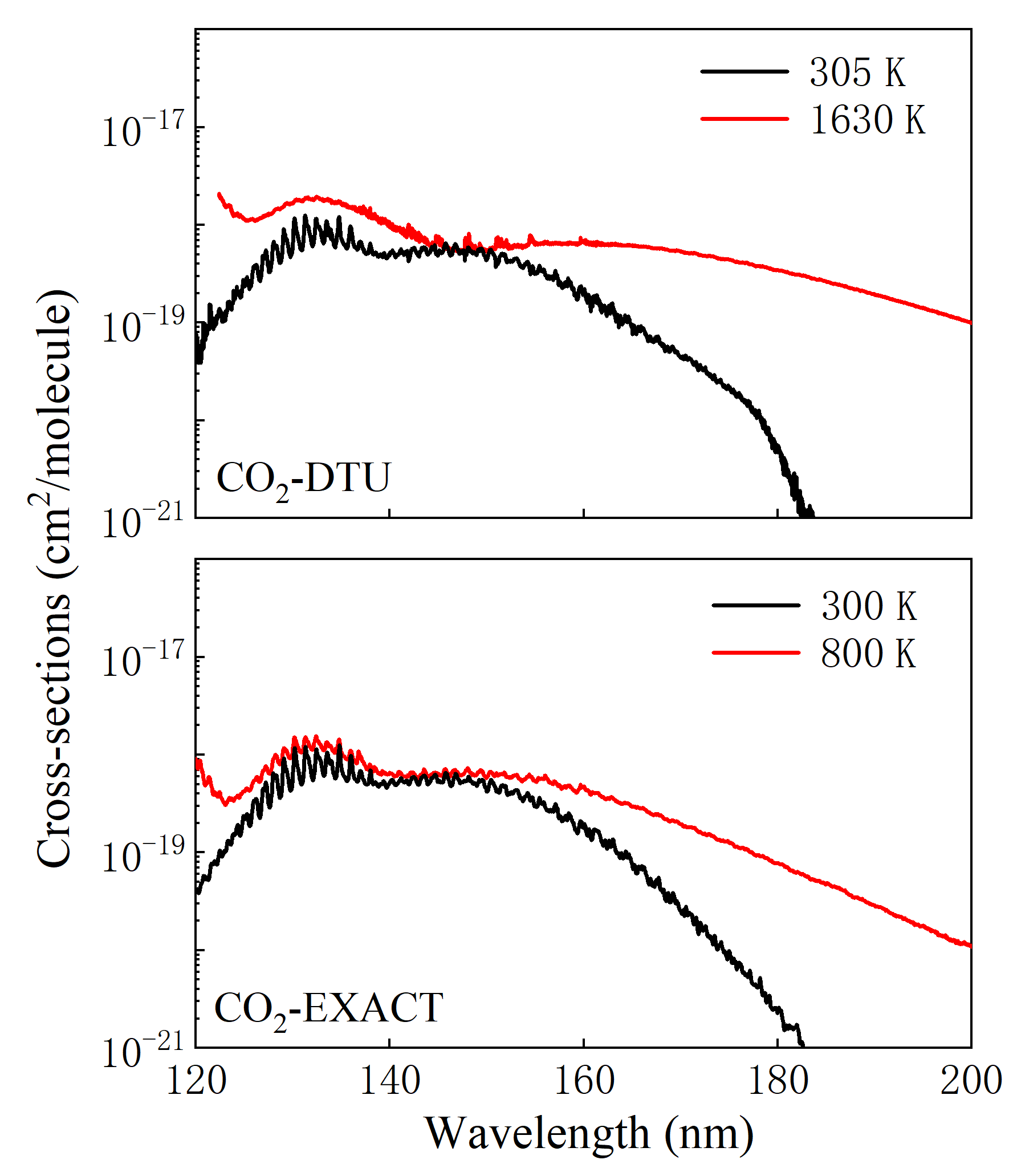} 
    \caption{CO\textsubscript{2}  spectrum overview from DTU and EXACT \citep{venot2018vuv} \citep{Fateev2021}} 
    \label{fig:CO2}
\end{figure}

Photodissociation cross sections for carbon dioxide (CO\textsubscript{2}) measured  by DTU and EXACT databases, see Fig.~\ref{fig:CO2}. Data from DTU span a wavelength range of 108.79--323.79~nm with a resolution of 0.01~nm, covering a temperature range from 305~K to 1630~K for pressures of 1 to 1.0647~bar. The measurements include cross sections at 305~K for wavelengths between 119.367~nm and 189.947~nm, at 550~K for wavelengths from 120.950~nm to 213.870~nm at 1.0647~bar, and at 1630~K for wavelengths extending from 122.420~nm up to 279.612~nm.

EXACT database also provides complementary data for a wavelength range of 114--230~nm with a resolution of 0.03~nm, covering temperatures from 150~K to 800~K under the pressure of 1 bar.

\subsubsection{\upshape H\textsubscript{2}O (DTU)\citep{Fateev2021}}

\begin{figure}
    \centering
    \includegraphics[width=0.9\textwidth]{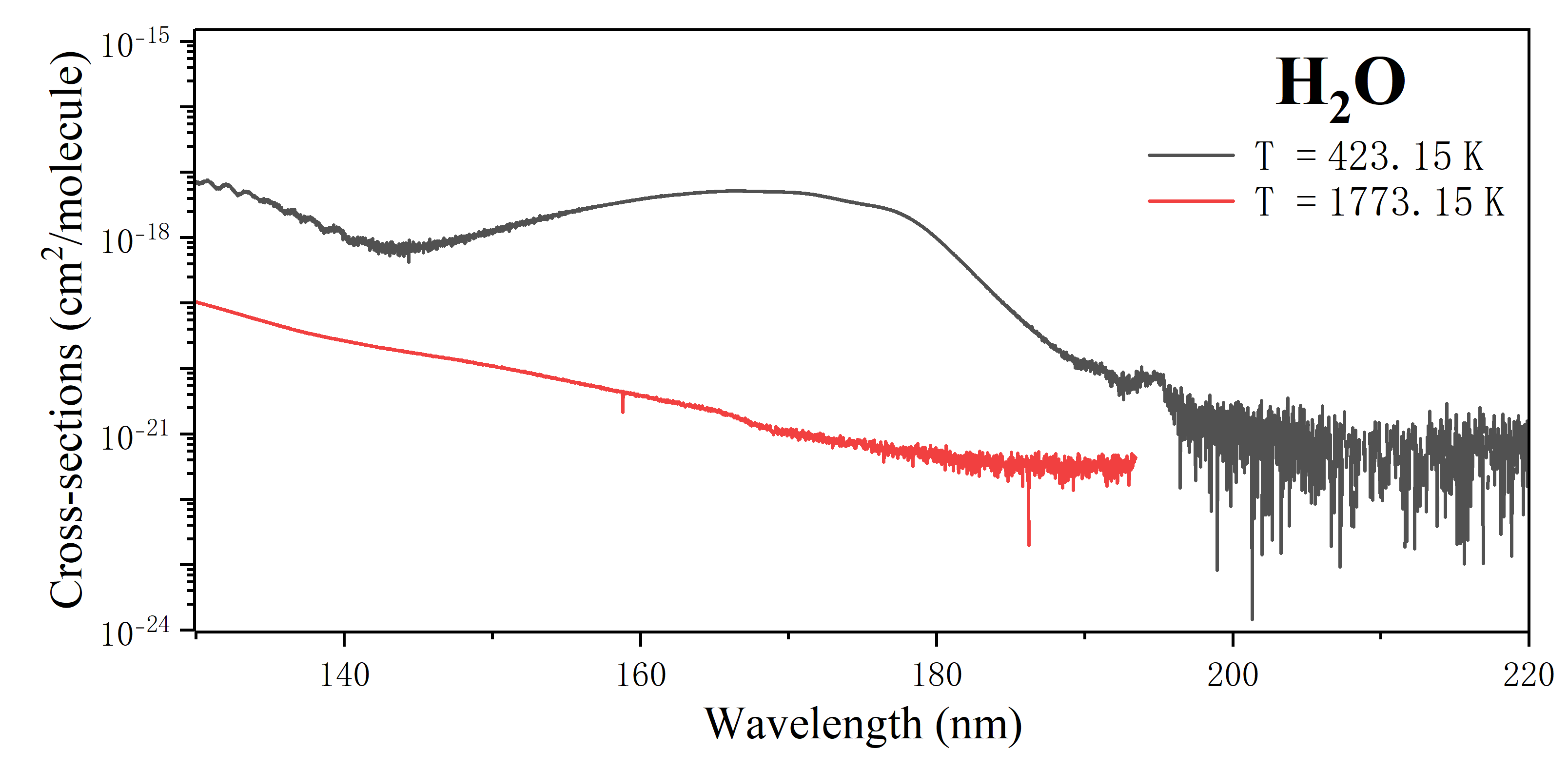} 
    \caption{H\textsubscript{2}O  spectrum overview \citep{Fateev2021}} 
    \label{fig:H2O}
\end{figure}

Photodissociation cross sections for water (H\textsubscript{2}O) were experimentally measured by DTU in the far-UV range. The cross sections span multiple wavelength ranges and resolutions, depending on the temperature. For temperatures of 423.15~K and 573.15~K, the wavelength range is 110--230~nm, with a resolution of 0.015~nm. At higher temperatures of 1630~K and 1773.15~K, the wavelength ranges extend to 108--237~nm and 182--231~nm, with a finer resolution of 0.010~nm under pressure of 1~bar, see Fig~\ref{fig:H2O}.

Water is a key molecule in many exoplanetary atmospheres and use of these cross sections has already shown important
effects in models of potentially habitable exoplanets \citep{20RaScHa.H2O,24BrScRa.H2O}.

\subsection{Larger Molecules}

\subsubsection{\upshape C\textsubscript{2}H\textsubscript{2} (EXACT) \citep{hcch}}

\begin{figure}
    \centering
    \includegraphics[width=0.9\textwidth]{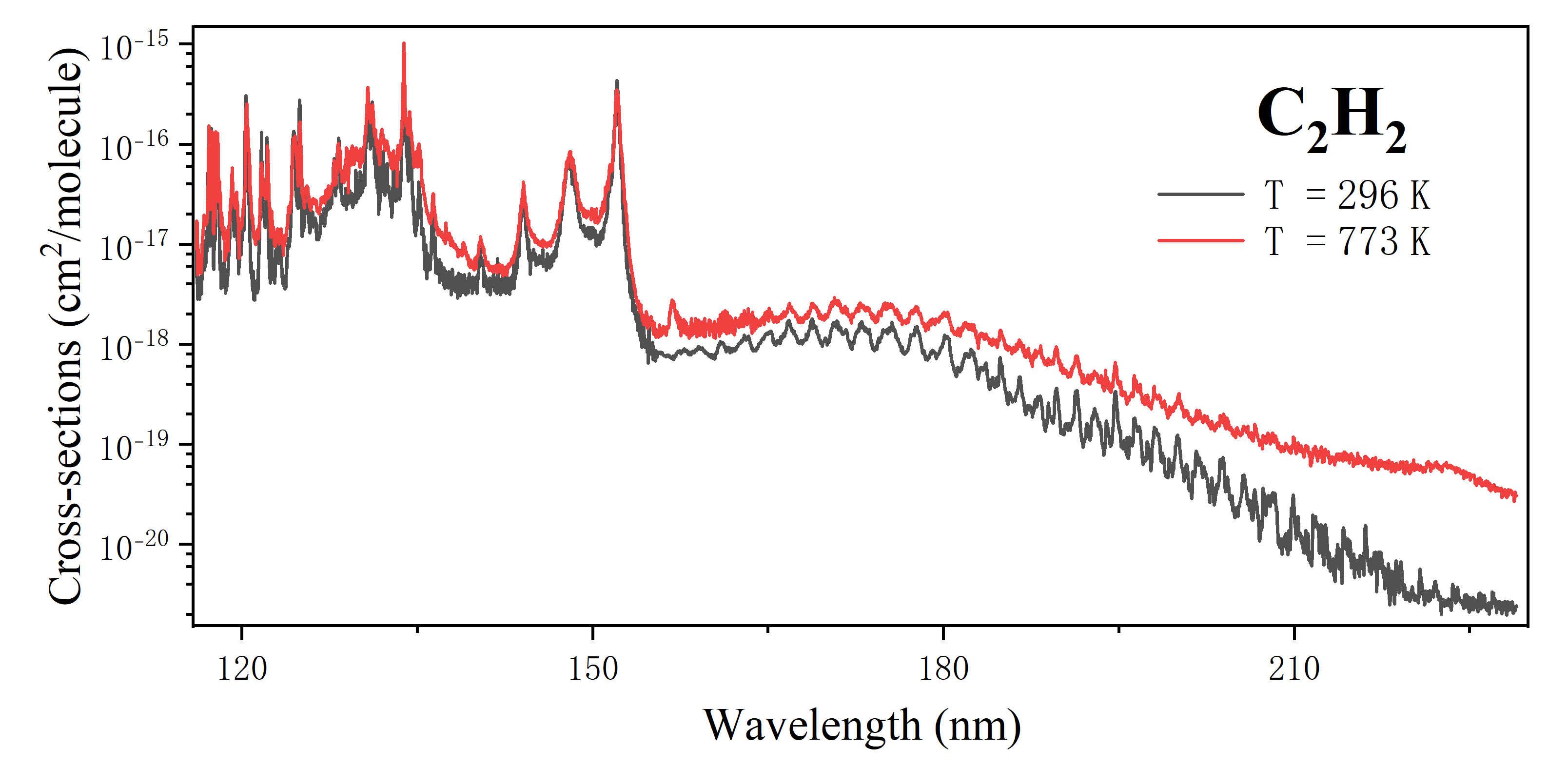} 
    \caption{C\textsubscript{2}H\textsubscript{2}  from EXACT spectrum overview \citep{hcch}} 
    \label{fig:C2H2}
\end{figure}

Photodissociation cross sections for acetylene (C\textsubscript{2}H\textsubscript{2}) were experimentally measured by Fleury and co-workers. The data span a wavelength range from 116 nm to 228 nm, with a resolution of 0.02 nm. Cross sections were provided for temperatures ranging from 296 K to 773 K under a pressure of 1 bar, see Fig.~\ref{fig:C2H2}. These cross sections were used to investigate their impact on exoplanet atmospheres  \citep{hcch}.

\subsubsection{\upshape H\textsubscript{2}CO (DTU) \citep{Fateev2021}}

\begin{figure}
    \centering
    \includegraphics[width=0.9\textwidth]{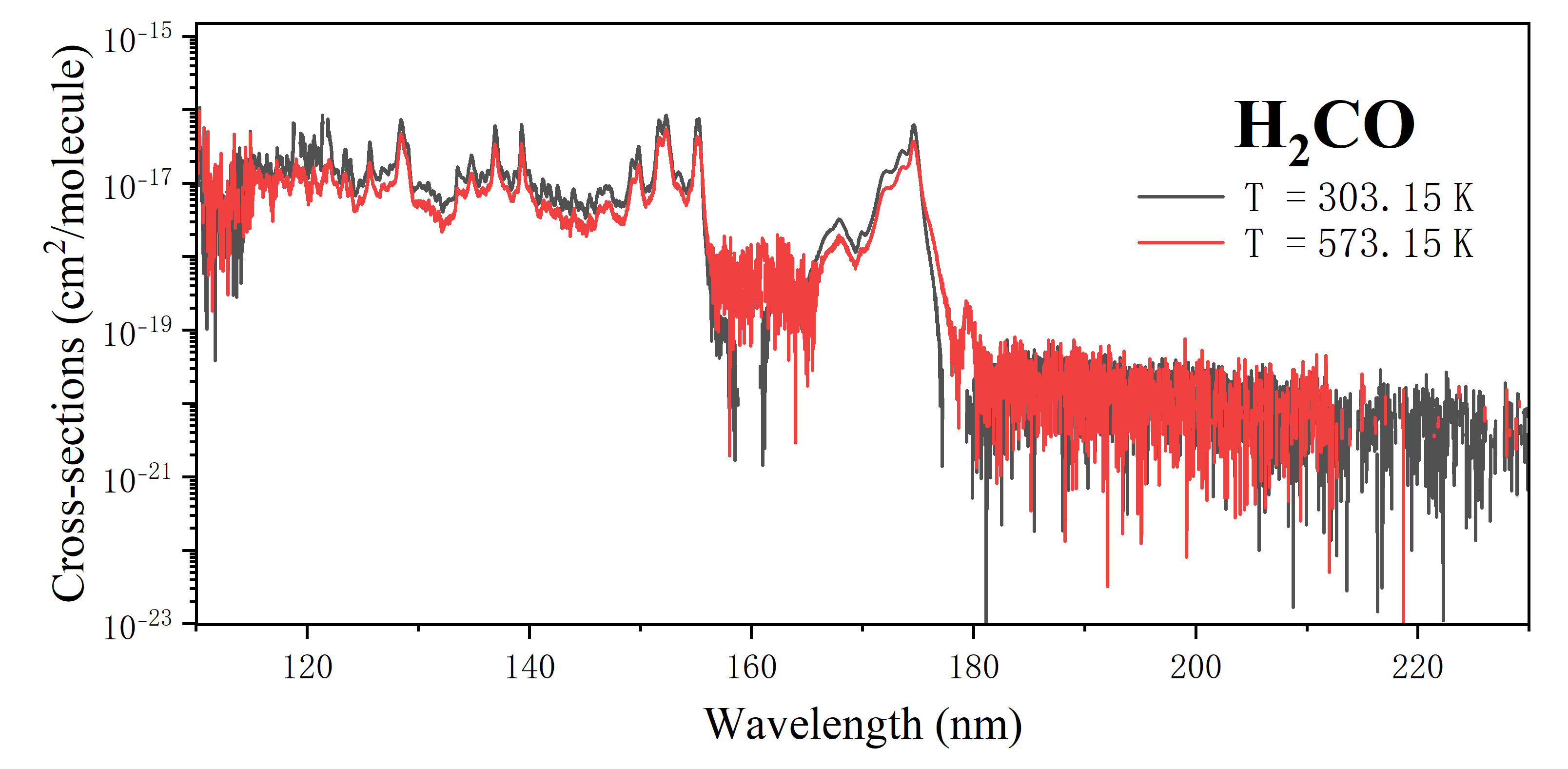} 
    \caption{H\textsubscript{2}CO  spectrum overview \citep{Fateev2021}} 
    \label{fig:H2CO}
\end{figure}

Photodissociation cross sections for formaldehyde (H\textsubscript{2}CO) were experimentally measured by DTU in the far-UV range. The data span a wavelength range from 110~nm to 230~nm with a resolution of 0.015~nm. Cross sections were provided for multiple temperatures, 303.15~K, 353.15~K, 423.15~K, and 573.15~K under pressure of 1~bar. They are illustrated in Fig.~\ref{fig:H2CO}.

\subsubsection{\upshape NH\textsubscript{3}  (DTU)\citep{Fateev2021}}

\begin{figure}
    \centering
    \includegraphics[width=0.9\textwidth]{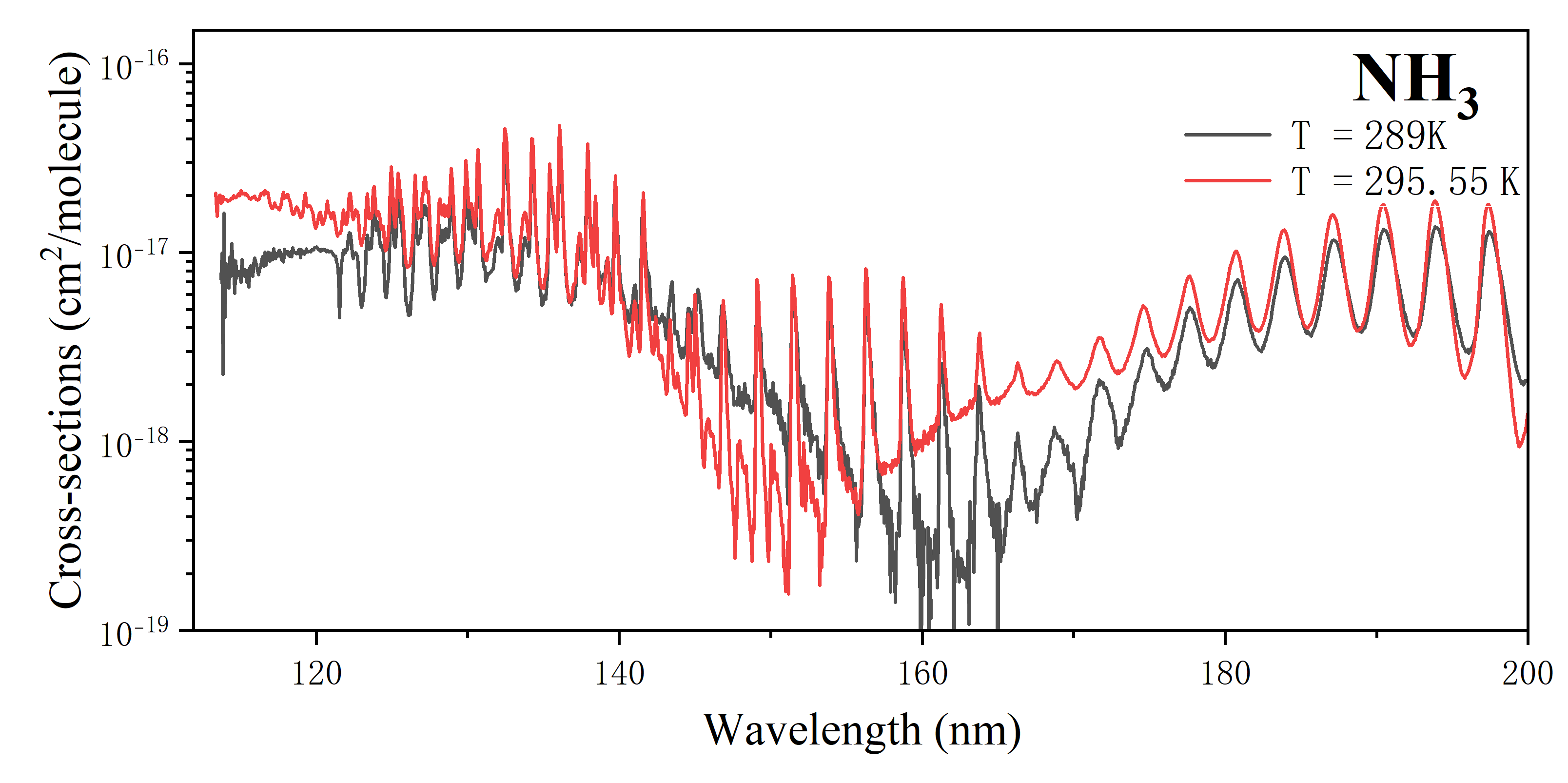} 
    \caption{NH\textsubscript{3}  spectrum overview \citep{Fateev2021}} 
    \label{fig:NH3}
\end{figure}

Photodissociation cross sections for ammonia (NH\textsubscript{3}) were measured experimentally by DTU in the far-UV range. The cross sections span a wavelength range from 113~nm to 202~nm, with a resolution of 0.015~nm. Data were provided for two temperatures, 289~K and 295.55~K under pressure of 1~bar, see Fig.~\ref{fig:NH3}.

Additional NH\textsubscript{3} cross sections up to 573~K will be published soon (Fleury et al., in prep.) and will be available via the \textsc{EXACT} database.

\subsubsection{\upshape C\textsubscript{2}H\textsubscript{4}  (DTU)\citep{Fateev2021}}

\begin{figure}
    \centering
    \includegraphics[width=0.9\textwidth]{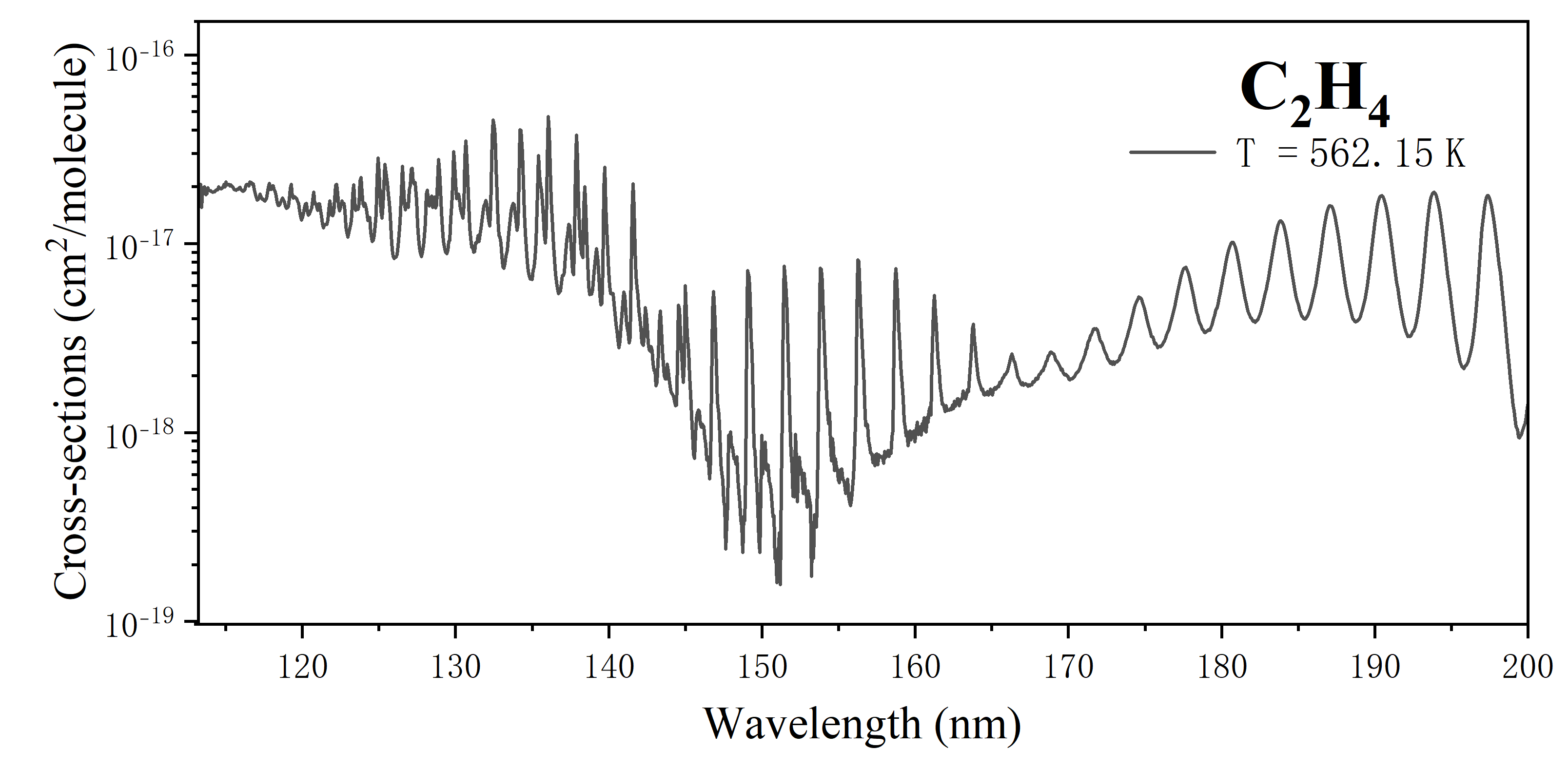} 
    \caption{C\textsubscript{2}H\textsubscript{4}  spectrum overview \citep{Fateev2021}} 
    \label{fig:C2H4}
\end{figure}

Photodissociation cross sections for ethylene (C\textsubscript{2}H\textsubscript{4}) were experimentally measured by DTU in the far-UV range. The data span a wavelength range from 113~nm to 201~nm, with a resolution of 0.065~nm. cross sections were provided for a single temperature of 562.15~K under pressure of 1~bar, see Fig.~\ref{fig:C2H4}.

\section{Database Structure}

Access to the database is implemented using the Django web framework \citep{django}, which is written in the Python programming language. The data structure of \textsc{ExoPhoto} is an extension of the original \textsc{ExoMol} data structure as anticipated by \cite{jt898}. This structure is designed to give a full description of the meta-data
for each file and in such a manner that it can be routinely used for both downloading and updating data employing an API (application program interface), see \citet{jt939}.
Table \ref{tb4} provides a specification summary  of  photodissociation data base. The file types constituting the data base are described in detail below.%

\begin{table}
\caption{\label{tb4} Specification of the \textsc{ExoPhoto} database file types. There is a single master file describing the recommended data for $N_{\rm mol}$ molecules which comprise $N_{\rm iso}$ per molecule.} 

\centering
\begin{tabular}{lcll}
\hline
\textbf{File extension} & \textbf{$N_{\rm files}$} & \textbf{File Type} & \textbf{Contents} \\ 
\hline 
.all & 1 & Master & \begin{tabular}[c]{@{}c@{}}Single file defining contents of the  \textsc{ExoPhoto} database\end{tabular} \\
.pdef.json & $N_{\rm tot}$ & Definition & \begin{tabular}[c]{@{}c@{}}Defines contents of other files for each isotopologue\end{tabular} \\
.model & $N_{\rm iso}$ & Model & Model specification for each isotopologue \\
.photo & $a$ & \begin{tabular}[c]{@{}c@{}}Photodissociation cross sections\end{tabular} & \begin{tabular}[c]{@{}c@{}}Photodissociation temperature-dependent cross sections \\(wavelength, nm) in .photo format for each isotopologue\end{tabular} \\
\hline
\multicolumn{4}{l}{$N_{\rm mol}$: number of molecules in the database} \\ 
\multicolumn{4}{l}{$N_{\rm tot}$: total number of .json files, summed across all molecules} \\ 
\multicolumn{4}{l}{$N_{\rm iso}$: number of isotopologues considered for each molecule} \\ 
\multicolumn{4}{l}{There are $N_{\rm tot}$ sets of .json files, one for each isotopologue} \\ 
\multicolumn{4}{l}{$a$: number of temp-dependent isotopologues files summed over $N_{\rm tot}$} \\ 
\hline
\end{tabular}
\end{table}


\subsection{File Naming Convention}
 The file naming convention in the \textsc{ExoPhoto} database ensures unique, descriptive, and machine-readable file names for each molecular isotopologue. 

 Files with the extensions \texttt{.json} and \texttt{.model} include only the ``common part'', which consists of the isotopologue's Iso-slug followed by the dataset identifier (\textsc{dataset\_name}). The Iso-slug is a plain ASCII-text, XML-safe identifier unique to each isotopologue. For instance, the isotopologue \(\ce{^1H^{35}Cl}\) is represented by the Iso-slug \texttt{1H-35Cl}. Following the Iso-slug is the database resource identifier, indicated by the ``\textsc{File\_extensions}'' column in Table \ref{tb4}. This identifier reflects specific datasets associated with the isotopologue. For example, there are two independent  photodissociation cross section sets available for \(\ce{^1H^{35}Cl}\) which are distinguished by the identifiers  \texttt{PTY} and \texttt{PhoMol}. Thus, file names for different types of data associated with each isotopologue follow the general pattern: \texttt{<iso\_slug>\_\_<dataset\_name>}, such as \texttt{1H-35Cl\_\_PTY\_\_}. The dataset identifiers currently in use within \texttt{ExoPHOTO} include: \texttt{PTY} and \texttt{MYTHOS} for data from \texttt{ExoMol}, \texttt{PhoMol},  \texttt{UGAMOP}, \texttt{DTU}, and \texttt{EXACT}.

For photodissociation cross section files with the extension .photo, additional information is appended to the common part to indicate specific details of the dataset. The .photo file names contain a wavelength range (nm), representing the full coverage of the photodissociation process, followed by the temperature in Kelvin (K), pressure in bar and the binning interval (nm), for which the cross sections have been computed, after the common part. The naming structure for these files is as follows:
\texttt{<iso\_slug>\_\_<dataset\_name>\_\_[MINwavelength]-[MAXwavelength]\_\_[Temperature]\_\_[Pressure]\_\_[Delta\_wavelength].photo}

For example, a file representing cross section data for \ce{^1H^{35}Cl} at a temperature of 100 K, pressure of 0 bar, with a wavelength range from 100 to 400  nm and a grid of 0.1 nm, would be named \texttt{1H-35Cl\_\_PTY\_\_100.0-400.0\_\_0100\_\_0\_\_0.1.photo}. We note that so far none of the calculated cross sections make any allowance for pressure effects and therefore provide the photodissociation cross sections at zero pressure; it is not expected that the photodissociation cross sections will
show significant pressure dependence.

\subsection{Photodissociation Cross Section File}

Photodissociation cross sections are stored in \textsc{ExoPhoto} files with the extension \texttt{.photo}. The primary columns include the wavelength (column 1, $\lambda$, in nm) and the total photodissociation cross section (column 2, $\sigma$, in cm$^2$/molecule). Additional columns, if applicable, provide partial cross sections for individual final electronic states of the system (e.g., $2\ ^1\Pi$, $A'\ ^1\Sigma^+$). The fields are separated by a single space and
the partial cross sections sum to give the total cross section.

The photodissociaition  cross sections provided by \textsc{ExoPhoto} are temperature- and in some cases, pressure-dependent. Each \texttt{.photo} file provides  data for a single combination of temperature and pressure. 
The corresponding temperatures and pressures are indicated in the file header, which specifies the temperature (in K) and the pressure (in bar) under which the cross sections were determined.  

The Fortran format for the complete \texttt{.photo} file is defined as \texttt{(F12.6, 1x, ES14.8)}, where each cross section field is separated by a single space. This file structure allows for representation of wavelength-dependent  photodissociation data for various isotopologues, with all other information encoded in the file names and the directory structure.

\subsubsection{Two-column Data Set}

For most molecules, only two columns are provided: wavelength and total cross section, see Table \ref{t:structure}.

\begin{table}
  \caption{A general two-column structure of the \texttt{.photo} file.}
  \label{t:structure}
  \centering
  \begin{tabular}{@{}llll@{}}
    \toprule
    \textbf{Field}
      & \textbf{Fortran Format}
      & \textbf{C Format}
      & \textbf{Description} \\
    \midrule
    Wavelength ($\lambda$)
      & \texttt{F12.6}
      & \texttt{\%12.6f}
      & Wavelength, in nm \\
    Total ($\sigma$)
      & \texttt{ES14.8}
      & \texttt{\%14.8e}
      & Total cross section, in cm$^2$/molecule \\
    \bottomrule
  \end{tabular}
\end{table}

\subsubsection{Multiple Column Data Set}
However, as noted above, for certain molecules such as \ce{CS} and \ce{MgH}, the cross section data are divided by outgoing channels, corresponding to different dissociating final states. For example, the \texttt{.photo} file format of CS molecule is shown in Table \ref{format-p2}.

\begin{table}
  \caption{Example of a multi‐column structure of the \texttt{.photo} file in \textsc{ExoPhoto} for \ce{CS}, containing total as well as partial cross sections for individual final electronic states of the system.}
  \label{format-p2}
  \centering
  \begin{tabular}{@{}llll@{}}
    \toprule
    \textbf{Field}
      & \textbf{Fortran Format}
      & \textbf{C Format}
      & \textbf{Description} \\
    \midrule
    Wavelength ($\lambda$)
      & \texttt{F12.6}
      & \texttt{\%12.6f}
      & Wavelength, in nm \\
    Total ($\sigma$)
      & \texttt{ES14.8}
      & \texttt{\%14.8e}
      & Total cross section, in cm$^2$/molecule \\
    $2\,^1\Pi$
      & \texttt{ES14.8}
      & \texttt{\%14.8e}
      & Cross section for the $2\, ^1\Pi$ state, in cm$^2$/molecule \\
    $A'\,^1\Sigma^+$
      & \texttt{ES14.8}
      & \texttt{\%14.8e}
      & Cross section for the $A'\, ^1\Sigma^+$ state, in cm$^2$/molecule \\
    $3\,^1\Pi$
      & \texttt{ES14.8}
      & \texttt{\%14.8e}
      & Cross section for the $3\, ^1\Pi$ state, in cm$^2$/molecule \\
    $B'\,^1\Sigma^+$
      & \texttt{ES14.8}
      & \texttt{\%14.8e}
      & Cross section for the $B'\,^1\Sigma^+$ state, in cm$^2$/molecule \\
    $4\,^1\Pi$
      & \texttt{ES14.8}
      & \texttt{\%14.8e}
      & Cross section for the $4\, ^1\Pi$ state, in cm$^2$/molecule \\
    $A\,^1\Pi$
      & \texttt{ES14.8}
      & \texttt{\%14.8e}
      & Cross section for the $A\, ^1\Pi$ state, in cm$^2$/molecule \\
    \bottomrule
  \end{tabular}
\end{table}

\subsection{The Definition File}
The core information about each isotopologue is contained within its JSON definition file. JSON (JavaScript Object Notation) is a lightweight, human-readable data-interchange format commonly used for organizing and transmitting structured data \citep{json}.

In this case, the \texttt{.pdef.json} definition file adheres to the \textsc{ExoPhoto} format \texttt{<iso\_slug>\_\_<dataset\_name>.pdef.json}. This file specifies the data available from \textsc{ExoPhoto} for a given isotopologue and describes potential applications of the data.  

Within this format, information provided is grouped into several main sections (highlighted in bold in Table \ref{format}), with each section describing a specific characteristic or property of the molecule. Each main section may contain multiple subsections that further detail aspects of that category.

For some isotopologues, the JSON files also contain information about different photodissociation channels, providing data on branching ratios or quantum yields. In photodissociation studies, according to \citet{branchingratio} and \citet{Quantumyield}, a channel refers to a specific dissociation pathway characterized by distinct photofragments or electronic states. The branching ratio describes how the total photodissociation yield is distributed among various channels, effectively indicating the fraction of dissociation events occurring via each specific channel. By contrast, the quantum yield provides a measure of the overall efficiency of the photodissociation process, defined as the total number of dissociation events per photon absorbed, regardless of the channel involved. Thus, branching ratios detail the internal distribution among dissociative channels, while quantum yields quantify the global efficiency of photodissociation.

\begin{table}
  \caption{Definition of the .pdef.json file format; each entry starts on a new line.}
  \label{format}
  \centering
  \renewcommand{\arraystretch}{1.3}
  \begin{tabular}{>{\bfseries}l c c p{8cm}}
    \hline
    \textbf{Field}
      & \textbf{Fortran Format}
      & \textbf{C Format}
      & \textbf{Description} \\
    \hline
    \multicolumn{4}{l}{\textbf{Header Information}} \\
    \texttt{iso\_formula}
      & \texttt{A27}
      & \texttt{\%27s}
      & Isotopologue chemical formula \\
    \texttt{iso\_slug}
      & \texttt{A160}
      & \texttt{\%160s}
      & Isotopologue identifier (constructed from isotopic composition and element symbol) \\
    \texttt{inchikey}
      & \texttt{A27}
      & \texttt{\%27s}
      & InChIKey identifier for the molecule \\
    \texttt{inchi}
      & \texttt{A20}
      & \texttt{\%20s}
      & Standard InChI representation for the molecule \\
    \texttt{mass\_in\_Da}
      & \texttt{F12.6}
      & \texttt{\%12.6f}
      & Molecular mass in Daltons (Da) \\

    \multicolumn{4}{l}{\textbf{Dataset Information}} \\
    \texttt{name}
      & \texttt{A10}
      & \texttt{\%10s}
      & Name of the resource dataset \\
    \texttt{version}
      & \texttt{I8}
      & \texttt{\%8d}
      & Version of the dataset (format: YYYYMMDD) \\
    \texttt{max\_temperature}
      & \texttt{F8.2}
      & \texttt{\%8.2f}
      & Maximum temperature in dataset for photodissociation cross sections \\
    \texttt{min\_wavelength}
      & \texttt{F8.2}
      & \texttt{\%8.2f}
      & Minimum wavelength in dataset (nm) \\
    \texttt{max\_wavelength}
      & \texttt{F8.2}
      & \texttt{\%8.2f}
      & Maximum wavelength in dataset (nm) \\

    \multicolumn{4}{l}{\textbf{Units}} \\
    \texttt{units.T}
      & \texttt{A2}
      & \texttt{\%2s}
      & Temperature unit (K) \\
    \texttt{units.p}
      & \texttt{A3}
      & \texttt{\%3s}
      & Pressure unit (bar) \\
    \texttt{units.w}
      & \texttt{A3}
      & \texttt{\%3s}
      & Wavelength unit (nm) \\

    \multicolumn{4}{l}{\textbf{Columns Information}} \\
    \texttt{Wavelength}
      & \texttt{A10}
      & \texttt{\%10s}
      & First column: Wavelength (nm) \\
    \texttt{Total}
      & \texttt{A10}
      & \texttt{\%10s}
      & Second column: Total photodissociation cross section (cm$^2$/molecule) \\
    \texttt{Channel}
      & \texttt{A10}
      & \texttt{\%10s}
      & Subsequent columns: cross section per channel (cm$^2$/molecule) \\

    \multicolumn{4}{l}{\textbf{File Information}} \\
    \texttt{files.p}
      & \texttt{I3}
      & \texttt{\%3d}
      & Pressure (bar) \\
    \texttt{files.T}
      & \texttt{F6.2}
      & \texttt{\%6.2f}
      & Temperature (K) \\
    \texttt{files.delta\_wavelength}
      & \texttt{F4.1}
      & \texttt{\%4.1f}
      & Wavelength interval (nm) \\
    \texttt{files.nlines}
      & \texttt{I6}
      & \texttt{\%6d}
      & Number of lines in the file \\
    \texttt{files.filename}
      & \texttt{A40}
      & \texttt{\%40s}
      & Filename for photodissociation cross section file (constructed from parameters) \\
    \hline
  \end{tabular}
\end{table}

Appendix \ref{appendix:json_example} presents the definition file \texttt{.pdef.json} for the  $^{12}$C$^{32}$S molecule  from the photodissociation dataset UGAMOP-CS, as a typical example. These standardized fields allow consistent identification and reference across various isotopologues.

The definition file serves following purposes.

\subsubsection{Standardized \textsc{ExoPhoto} File Usage}

The dataset format used in this database follows \textsc{ExoMol} standards, designed to provide comprehensive photodissociation cross section data for a given isotopologue. Each isotopologue is described by fields such as the isotopologue formula (\texttt{iso\_formula}), the simplified slug name (\texttt{iso\_slug}), InChIKey (\texttt{inchikey}), InChI string (\texttt{inchi}), and molecular mass in Daltons (\texttt{mass\_in\_Da}). These standardized fields allow consistent identification and reference across various isotopologues.

The InChI encodes detailed structural information, such as the connectivity of atoms and the presence of hydrogen. The InChIKey, a hashed version of the InChI, is a compact, fixed-length string designed for easier indexing and searching. The process of generating an InChIKey involves applying a hashing algorithm to the full InChI string. This algorithm compresses the detailed structural information into a unique 27-character alphanumeric code, divided into three blocks separated by hyphens: the first block encodes the connectivity, the second encodes additional structural features, and the third indicates the version and checksum of the key. Both identifiers ensure consistency and compatibility across databases \citep{inchi}.

For the general species HCl, the InChI represents the molecule as a collection of isotopologues, defined by the averaged molecular weight of 36.46~amu. This molecular weight reflects the natural abundance of its isotopes \( ^1\text{H}, ^2\text{H}, ^{35}\text{Cl}, \text{and} ^{37}\text{Cl} \). The InChI for this ``average'' HCl is \texttt{1S/ClH/h1H}, and the corresponding InChIKey is \texttt{VEXZGXHMUGYJMC-UHFFFAOYSA-N}. However, this representation does not account for specific isotopic compositions. To distinguish specific isotopologues, an isotopic layer is added to the InChI string \citep{nist_inchi}.

The four major isotopologues of HCl are \(^1\text{H}{}^{35}\text{Cl}\), \(^1\text{H}{}^{37}\text{Cl}\), \(^2\text{H}{}^{35}\text{Cl}\) (DCl), and \(^2\text{H}{}^{37}\text{Cl}\). Each isotopologue has a unique InChI and InChIKey, reflecting its specific isotopic composition. Table \ref{tab:inchi_hcl}  summarizes the InChI and InChIKey representations for both general HCl and its specific isotopologues as illustrated  in Table~\ref{tab:inchi_hcl}. 

\begin{table}
\centering
\caption{InChI and InChIKey for General HCl and Specific Isotopologues \citep{nist_inchi,inchi_softwares}}
\label{tab:inchi_hcl}
\renewcommand{\arraystretch}{1.3}
\begin{tabular}{p{0.2\linewidth} p{0.2\linewidth} p{0.25\linewidth}}
\hline
\textbf{Description} & \textbf{InChI} & \textbf{InChIKey} \\
\hline
General HCl (average molecular weight: 36.46 amu)& \texttt{1S/ClH/h1H} & \texttt{VEXZGXHMUGYJMC-UHFFFAOYSA-N} \\
\(^1\text{H}-{}^{35}\text{Cl}\) & \texttt{1S/ClH/h1H/i1+0} & \texttt{VEXZGXHMUGYJMC-UHFFFAOYSA-N} \\
\(^1\text{H}-{}^{37}\text{Cl}\) & \texttt{1S/ClH/h1H/i1+0;2+0} & \texttt{VEXZGXHMUGYJMC-NJFSPNSNSA-N} \\
\(^2\text{H}-{}^{35}\text{Cl}\) (DCl) & \texttt{1S/ClH/h1H/i1+1} & \texttt{VEXZGXHMUGYJMC-DYCDLGHISA-N} \\
\(^2\text{H}-{}^{37}\text{Cl}\) & \texttt{1S/ClH/h1H/i1+1;2+0} & \texttt{VEXZGXHMUGYJMC-DQGQKLTASA-N} \\
\hline
\end{tabular}
\end{table}

While InChIKeys are highly standardized, they are not always  unique in 
 cases, such as stereoisomers or isotopic variations 
   \citep{inchi,inchi_softwares}. This is why additional fields like the \texttt{isotopologue slug} (\texttt{iso\_slug}) or \texttt{molecular mass in Daltons} (\texttt{mass\_in\_Da}) are included for more precise identification in databases. 

\subsubsection{Improved Database Functionality}

The dataset includes photodissociation cross section (\texttt{photodissociation\_xsecs}) data, with detailed fields for maximum temperature (\texttt{max\_temperature}), wavelength range (\texttt{min\_wavelength}, \texttt{max\_wavelength}), and units for key parameters. This information facilitates database operations like sorting, filtering, and selecting based on temperature, pressure, and wavelength increment values, as stored in the \texttt{files} section. Each entry in \texttt{files} specifies conditions such as pressure (\texttt{p}), temperature (\texttt{T}), and wavelength increment (\texttt{delta\_wavelength}), with filenames indicating all these parameters. 

\subsubsection{Facilitated Updates}

Each dataset is version-controlled, indicated by the \texttt{version} field in the date format \texttt{YYYYMMDD}. This versioning supports easy updating of isotopologue data, allowing users to check for newer versions of specific datasets and automatically download modified files. This feature ensures up-to-date data usage without requiring code modifications \citep{jt898}.

\subsection{Spectroscopic Model  File}

Each calculated dataset  in \textsc{ExoPhoto} is based on a well-developed and comprehensive spectroscopic model for the specified isotopologue; as far as possible in the  \textsc{ExoMol}/\textsc{ExoPhoto} database the actual data provided is supplemented by detailed specification of the spectroscopic model used.
This takes the form of reference to the original paper in all cases and provision of input files used and code specification for the calculations where possible. In practice, at present it is only  \textsc{ExoMol} data for which we are able to provide the full spectroscopic model. 



For the $^{1}$H$^{35}$Cl isotope of HCl, \textsc{ExoMol} provides spectroscopic data available on the website in the form of a corresponding model file. Appendix \ref{model} shows the \texttt{.model} file for the $^{1}$H$^{35}$Cl molecule, which also serves as an input file for the code \textsc{Duo} \citep{jt609}.

\subsection{The Master File}

A master summary file, named \texttt{exophoto.all.json}, consolidates the contents of the whole database. This file, which is  available at \url{www.exomol.com/exophoto.all.json},  provides a computer readable (JSON format) list of  recommended  data sets, including those for each isotopologue under the \textsc{ExoPhoto} structure. 
The \texttt{master} file provides a summary of the database's contents and ensures easy access to the current version number of each dataset, enabling users to efficiently track updates. The file begins with general statistics about the database, including the total number of molecules, isotopologues, and datasets. Specifically, this database currently contains 20 molecules, 28 isotopologues, and data are taken from a total of 5 projects/databases. 

For each molecule, the file includes its names, chemical formula, and the number of associated isotopologues. An example of the \textsc{ExoPhoto} master file for HCl is given in the Appendix C, illustrating the common uniform structure used for all molecules, with differences only in specific data values; isotopologue-specific details are also provided, including unique identifiers such as InChI keys, iso-slugs, isotopic formulas, dataset names, and version numbers (see Appendix~\ref{exophotomaster}). The format of the \texttt{ExoPhoto.master.json} file is illustrated in Table~\ref{masterformat}.



\begin{table}
\caption{Extract from the \textsc{ExoPhoto} Master file showing the overall header and information
contained for HCl}
\label{masterformat}
\centering
\renewcommand{\arraystretch}{1.3}
\begin{tabular}{l l}
\hline
\textbf{Field} & \textbf{Description} \\
\hline
exophoto.master & ID \\
20241225 & Version number with format YYYYMMDD \\
20 & Number of molecules in the database \\
28 & Number of isotopologues in the database \\
5 & Number of datasets in the database \\
\textbf{Molecule Information} & \\
Molecule Name & The common name of the molecule (\texttt{HCl}) \\
Chemical Formula & The chemical formula of the molecule \\
Number of Isotopologues & Total number of isotopologues available for \texttt{HCl} \\
Databases & The databases where \texttt{HCl} isotopologues are listed \\
\textbf{Isotopologue Overview} & \\
\texttt{1H-35Cl} & Found in \texttt{PTY} and \texttt{PHOTO-PhoMol\_HCl} databases \\
\texttt{2H-35Cl} & Found only in the \texttt{PTY} database \\
\texttt{1H-37Cl} & Found only in the \texttt{PTY} database \\
\texttt{2H-37Cl} & Found only in the \texttt{PTY} database \\
\textbf{Isotopologue Details} & \\
Inchi Key & The InChIKey identifier for the isotopologue \\
Iso-slug & The isotopologue slug \\
IsoFormula & The isotopologue formula in a standardized format \\
Dataset Name & The dataset where the isotopologue data is available \\
Version Number & Version of the dataset containing isotopologue information \\
\hline
\end{tabular}
\end{table}

\section{Conclusions}
The \textsc{ExoPhoto} database is an extension of the \textsc{ExoMol} database, specifically developed to address the need for high-accuracy photodissociation cross section data for molecules of astronomical interest. \textsc{ExoPhoto} currently includes 12 molecules derived from theoretical models provided by three theoretical databases and 8 molecules supported by experimental data from including two from the experimental EXACT  databases.  These cross sections are clearly only a subset of those needed to model photon-driven chemistries at the top exoplanetary atmospheres and elsewhere. We are currently working on photodissociation for a number of diatomic molecules and expanding our treatment to triatomics with initial focus on  \ce{HCN} and \ce{H2S}.

An important future development of the \textsc{ExoPhoto} database will be the inclusions of non-local thermodynamic equilibrium (non-LTE) cross sections i.e. cross sections which are resolved according to the initial states of the target molecule. As the number of initial states can be very large, particularly for polyatomic molecules, this step will lead to a huge expansion in the size of the database and therefore needs to be approached with some care.
At present we are using OH as a prototype to understand how users will utilise non-LTE photodissociation data.
Initial state specified OH photodissociation cross sections suitable for non-LTE studies are available on request.

Currently those molecules for which branching ratios, where available, are given in terms of partial cross section for the outgoing electronic state. In
future we would aim to correlate these  outgoing final states  with specific states of the asymptotic fragments, for example diatomic states we would provide information on the atomic states formed. This will provide a more comprehensive description but is not entirely straightforward as the available partial cross sections contain some cross sections to states which actually only dissociate after undergoing a state crossing to another (currently unspecified) dissociative state. 

The DTU team aims to extend their datasets by photoabsorption cross section measurements of molecules relevant to various industrial applications (e.g., Carbon Capture Utilization and Storage (CCUS), \(\text{H}_2\) and biogas production, transportation, and storage). The molecules of interest are: HCl, Cl\(_2\), H\(_2\)S, NH\(_3\), CO\(_2\), CO, O\(_2\), H\(_2\)O, etc. The measurements will first be performed at ambient conditions (about 298~K and 1~atm), then at elevated pressures (at 298~K), and finally at elevated pressures and temperatures (up to 100~atm and 1300~K). The measurements will be conducted in the 110--240~nm spectral range using primary reference materials (gas mixtures).

\section*{Acknowledgements}

This work was supported by the European Research Council (ERC) under the European Union’s Horizon 2020 research and innovation programme via Advanced Grant 883830 and the STFC Project No. ST/Y001508/1. O. Venot acknowledges funding from Agence Nationale de la Recherche (ANR), project ``EXACT'' (ANR-21-CE49-0008-01) as well as from the Centre National d’\'{E}tudes Spatiales (CNES).

\section*{Data availability}

All the data discussed in this paper are available from the \textsc{ExoPhoto} website \url{https://exomol.com/exophoto/} 
except for non-LTE cross sections for OH which can be obtained on request from the corresponding
author.

\newpage
\appendix
\section{Example \texttt{pdef.json} File  by CS Molecule}
\label{appendix:json_example}

\begin{lstlisting}[language=json]
{
   ``isotopologue": {
        "iso_formula": "(12C)(32S)",
        "iso_slug": "12C-32S",
        "inchikey": "DXHPZXWIPWDXHJ-UHFFFAOYSA-N",
        "inchi": "1S/CS/c1-2",
        "mass_in_Da": 43.972071
    },
    "dataset": {
        "name": "UGAMOP-CS",
        "version": 20240621,
        "photodissociation_xsecs": {
            "max_temperature": 10000.0,
            "min_wavelength": 50.0,
            "max_wavelength": 5000.0,
            "units": {
                "T": "K",
                "p": "bar",
                "dwv": "nm"
            },
            "columns": [
                {"name": "wavelength", "units": "nm"},
                {"name": "Total", "units": "cm2/molecule"},
                {"name": "2(1PI)", "units": "cm2/molecule"},
                {"name": "A'(1SIGMA+)", "units": "cm2/molecule"},
                {"name": "3(1PI)", "units": "cm2/molecule"},
                {"name": "B'(1SIGMA+)", "units": "cm2/molecule"},
                {"name": "4(1PI)", "units": "cm2/molecule"},
                {"name": "A(1PI)", "units": "cm2/molecule"}
            ],
            "files": [
                {
                    "p": 0,
                    "T": 1000,
                    "dwv": 3.33,
                    "nlines": 1500,
                    "filename": "12C-32S__UGAMOP-CS__50.0-5000.0__1000__0__3.33.photo"
                },
                {
                    "p": 0,
                    "T": 2000,
                    "dwv": 3.33,
                    "nlines": 1500,
                    "filename": "12C-32S__UGAMOP-CS__50.0-5000.0__2000__0__3.33.photo"
                },
                
                ...
                
                {
                    "p": 0,
                    "T": 10000,
                    "dwv": 3.33,
                    "nlines": 1500,
                    "filename": "12C-32S__UGAMOP-CS__50.0-5000.0__10000__0__3.33.photo"
                }
            ]
        }
    }
}
\end{lstlisting}

\newpage
\section{Example \textsc{ExoMol} spectroscopic \texttt{.model} file for  \texorpdfstring{$^{1}$H$^{35}$Cl}{HCl}}
\label{model}

\begin{lstlisting}[language=json]
**Input Parameters for HCl**

atoms H Cl

molecule HCl

(Total number of states taken into account)
nstates 2

(Total angular momentum quantum - a value or an interval)
jrot 0 - 100 

memory 32 GB

(SOLUTIONMETHOD 5POINTDIFFERENCES)

(Defining the integration grid)
grid
  npoints        251    (odd)
  range  0.5,3.00
  type 0   (nsub)
  re 0.9
end 

symmetry Cs(M)

**Diagonalizer and Contraction Parameters**

DIAGONALIZER 
 SYEV   (SYEVR)
end

CONTRACTION
  vib
  vmax  200          
  enermax 150000
END

**Potential Energy Surfaces**

poten 1
name "X 1Sigma+"
symmetry +
type   grid
lambda 0
mult   1
units Angstroms Eh
values
  0.80000001192092896  0.38017997475145210     
  0.82090001180768013  0.33623301634462599     
  ...
end

poten 2
name "3 1Sigma+"
symmetry +
type   grid
lambda 0
mult   1
units Angstroms Eh
values
  0.800000011920929  0.807352361388752
  0.82090001180768   0.76269762315556
  ...
end

**Intensity Parameters**

INTENSITY  
  absorption
  thresh_intes  1e-30
  thresh_line   1e-30
  temperature   10000 
  nspin 0.5  1.5
  linelist   3s-x-L3.00-J120 
  J,  0.0 - 120
  freq-window        0.0, 1000000.0
  energy low  0., 37239.9374, upper    80000, 1.00E+06 
end

**Dipole Moments**

dipole  1 2
name "<3S+ |DMZ|  X1Sigma+ >"
spin   0.0 0.0
lambda  0  0
type   grid
factor   1   (0, 1 or i)
units Angstroms ea0
values
  0.8000 	0.4966
  1.0000 	0.5004
  1.2750 	0.4411
  ...
end
\end{lstlisting}

\newpage
\section{Extract from the \textsc{ExoPhoto} Master file showing the specifications for the HCl isotopolgues.}
\label{exophotomaster}

\begin{lstlisting}[language=json]
{
  "ExoPhoto": {
    "ID": "exophoto.master.json",
    "version": "20241225"
  },
  "molecules": {
    "HCl": {
      "name": "Hydrogen chloride",
      "ordinary_formula": "HCl",
      "isotopologues": {
        "1H-35Cl": {
          "inchikey": "VEXZGXHMUGYJMC-UHFFFAOYSA-N",
          "iso_formula": "(1H)(35Cl)",
          "datasets": {
            "PTY": {
              "version": 20240916,
              "min_wavelength": { "value": 100, "units": "nm" },
              "max_wavelength": { "value": 400, "units": "nm" },
              "min_temperature": { "value": 0.01, "units": "K" },
              "max_temperature": { "value": 10000, "units": "K" },
              "nfiles": 34
            },
            "PhoMol": {
              "version": 20241001,
              "min_wavelength": { "value": 50, "units": "nm" },
              "max_wavelength": { "value": 500, "units": "nm" },
              "min_temperature": { "value": 0.01, "units": "K" },
              "max_temperature": { "value": 10000, "units": "K" },
              "nfiles": 34
            }
          }
        },
        "2H-35Cl": {
          "inchikey": "VEXZGXHMUGYJMC-UHFFFAOYSA-N",
          "iso_formula": "(2H)(35Cl)",
          "datasets": {
            "PTY": {
              "version": 20240916,
              "min_wavelength": { "value": 100, "units": "nm" },
              "max_wavelength": { "value": 400, "units": "nm" },
              "min_temperature": { "value": 0.01, "units": "K" },
              "max_temperature": { "value": 10000, "units": "K" },
              "nfiles": 34
            }
          }
        }
        // ... other HCl isotopologues ...
      }
    },
    "CS": {
      "name": "Carbon monosulfide",
      "ordinary_formula": "CS",
      "isotopologues": {
        "12C-32S": {
          "inchikey": "UEGCPOVBQZRIBG-UHFFFAOYSA-N",
          "iso_formula": "(12C)(32S)",
          "datasets": {
            "UGAMOP": {
              "version": 20240215,
              "min_wavelength": { "value": 50, "units": "nm" },
              "max_wavelength": { "value": 5000, "units": "nm" },
              "min_temperature": { "value": 1000, "units": "K" },
              "max_temperature": { "value": 10000, "units": "K" },
              "nfiles": 10
            }
          }
        }
      }
    }
    // ... other molecules ...
  }
}

\end{lstlisting}



\bibliographystyle{rasti}







\bsp	
\label{lastpage}
\end{document}